\documentclass[twocolumn]{aastex}
\fullcollaborationName{The Friends of AASTeX Collaboration}

\begin{document}


\title{Solar energetic particle propagation in wave turbulence and the possibility of wave generation}


\author{R.D. Strauss\altaffilmark{1}}
\affil{Center for Space Research, North-West University, Potchefstroom, 2522, South Africa}
\affil{National Institute for Theoretical Physics (NITheP), Gauteng, South Africa}

\and

\author{J.A. le Roux}
\affil{Center for Space Plasma and Aeronomic Research, University of Alabama in Huntsville, Huntsville, AL 3585, USA}
\affil{Department of Space Science, University of Alabama in Huntsville, Huntsville, AL 35899, USA}


\altaffiltext{1}{dutoit.strauss@nwu.ac.za}

\begin{abstract}

A complete theory for the complex interaction between solar energetic particles and the turbulent interplanetary magnetic field remains elusive. In this work we aim to contribute towards such a theory by modelling the propagation of solar energetic particle electrons in plasma wave turbulence. We specify a background turbulence spectrum, as constrained through observations, calculate the transport coefficients from first principles, and simulate the propagation of these electrons in the inner heliosphere. We have also, for the first time, included dynamical effects into the perpendicular diffusion coefficient. We show that such a ``physics-first" approach can lead to reasonable results, when compared qualitatively to observations. In addition, we include the effect of wave growth/damping due to streaming electrons and show that these particles can significantly alter the turbulence levels close to the Sun for the largest events. 

\end{abstract}




\section{Introduction}

During transient solar phenomena, such as solar flares and/or coronal mass ejections, high energy solar energetic particles (SEPs) are accelerated \citep[e.g.][]{remaesreview,remaesreview2}, where after these particles propagate in the turbulent heliospheric magnetic field to reach Earth and the inner heliosphere. A complete description of the interaction (i.e. scattering) between particles and these turbulent fields remains elusive. One the one hand, more observations (of both SEP particles, but also of the turbulent magnetic fields) near the Sun are needed, while, on the other hand, more detailed modelling of SEP propagation is needed on a more fundamental level. We advocate a so-called ``physics-first" approach in modelling SEPs: specify the background turbulence quantities as informed by observations and/or simulations and calculate the relevant transport quantities (i.e. diffusion coefficients) by using a realistic turbulence spectrum as input. Finally, a comparison between modelled and SEP intensities can be performed. To some extent this is a continuation of our effort described in \citet{straussetal2017}. This approach is supplemented by empirical modelling where the transport coefficients are treated as free parameters and fitted to particle observations \citep[e.g.][]{drogeteal2016}.

In this study we perform further refinements of our model and again apply it to the transport of impulsively accelerated, near-relativistic ($\sim 100$ keV) electrons. {We therefore focus on so-called {\it electron rich} events, where the SEP proton component can, for the most part, be neglected.} We calculate the transport coefficients for dynamical turbulence and also include previous neglected effects such as those that arise from a non-vanishing magnetic and cross helicity. We also attempt to quantify the background turbulence in more detail, specifically the wave-number at which the dissipation of turbulence starts being significant. In addition to the background (or rather, omni-present) solar wind turbulence, we also investigate whether SEP electrons can modify the background turbulence via the streaming instability: can a SEP event amplify/damp solar wind turbulence? Although this has been modelled and observed before for proton events \citep[e.g.][]{ngetal1999,desaietal2012}, we are unaware of any such simulations for electrons. Moreover, such a study is additionally motivated by the recent findings of \citet{aguedalario} that the derived mean-free-path for some events may depend on the intensity of the SEP events; an indication that more particles can lead to enhanced levels of scattering and hence a decrease in the mean-free-path.

\section{Solar wind fluctuations}

Magnetic turbulence is usually described by Reynolds decomposing the magnetic field, $\vec{B}$, into a locally uniform background field, $\vec{B_0} $, and a random turbulent component, $\vec{b}(x,y,z)$, such that

\begin{equation}
\vec{B} = \vec{B_0} + \vec{b}(x,y,z).
\end{equation}

Averaging over long time periods, yields $\langle \vec{B} \rangle= \vec{B_0}$, while $ \delta B^2=\langle \vec{b}^{2} \rangle$ is the variance of the fluctuating component. Furthermore, we assume the turbulence to be transversal, i.e. $\vec{B_0} \cdot \vec{b} = 0$ (this is equivalent to assuming a strong guide field, i.e. if $\vec{B_0} = B_0 \mathbf{z}$, then the $\mathbf{z}$ component of $\vec{b}$ is negligible, $\vec{b} \cdot \mathbf{z} \ll B_0$) and that the energy contained in the turbulence is much less than that of the background field, $\delta B^2 \ll B_0^2$.


\begin{figure*}
\begin{center}
\includegraphics[width=80mm]{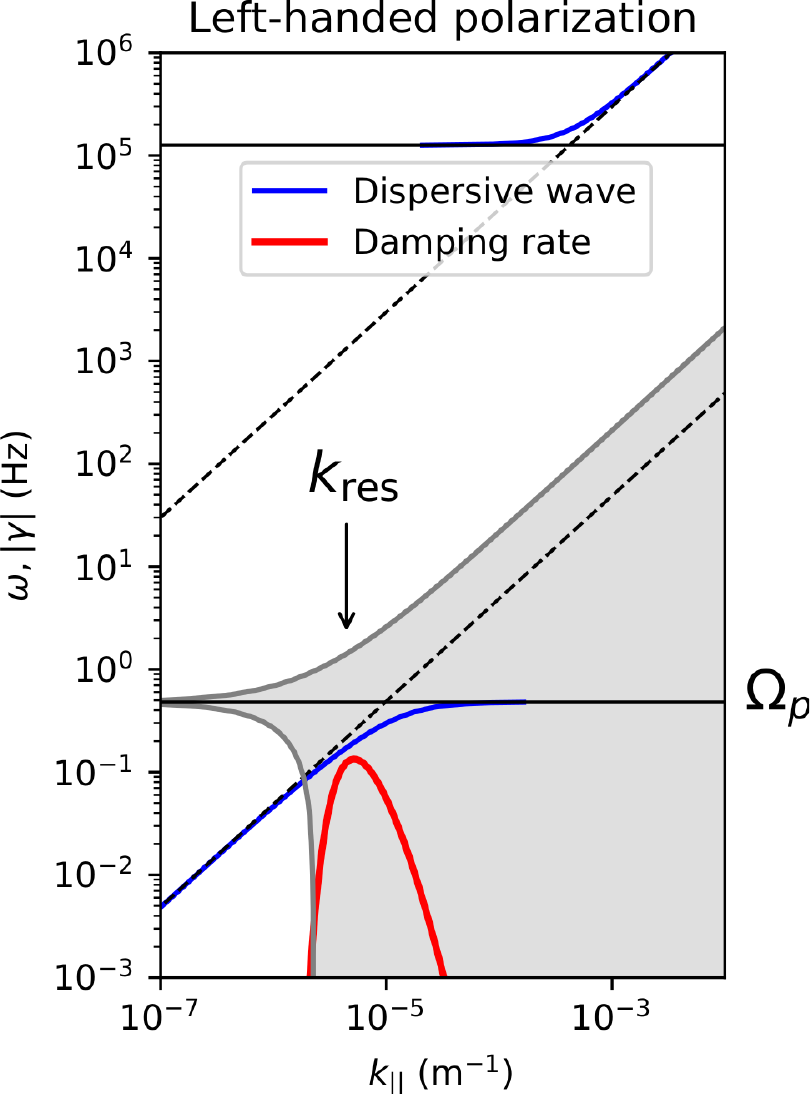}
\hspace{5mm}
\includegraphics[width=80mm]{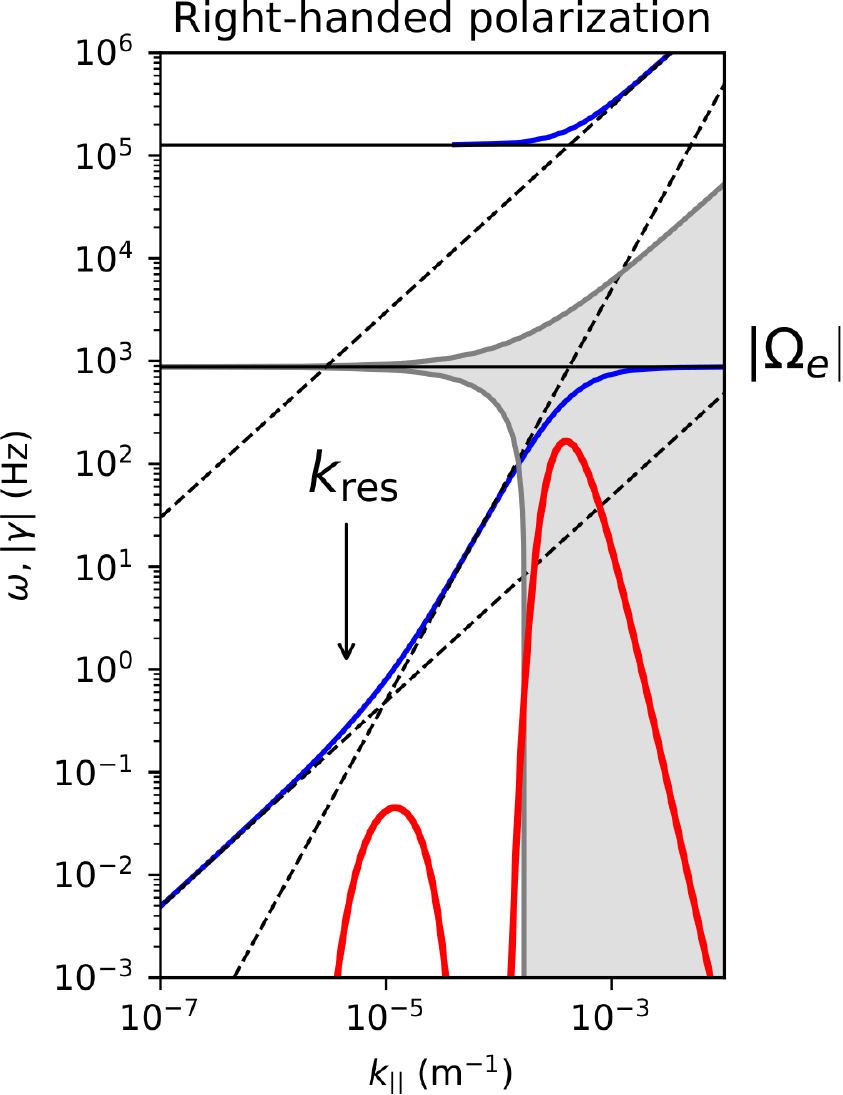}
\caption{The figure shows dispersion diagrams (solid blue curves) for LH polarized (left panel) and RH polarized (right panel) parallel propagating waves in the cold plasma limit. The solid red curves are the approximate cyclotron damping rates (valid for small damping rates). All of these quantities are given in Appendix \ref{Sec:appendix_2}. The great shaded areas indicate regions where thermal particles are believed to gyro-resonate very effectively with the the different wave modes.  \label{fig:dispersion_diagram}}
\end{center}
\end{figure*}

\begin{figure*}
\begin{center}
\includegraphics[width=160mm]{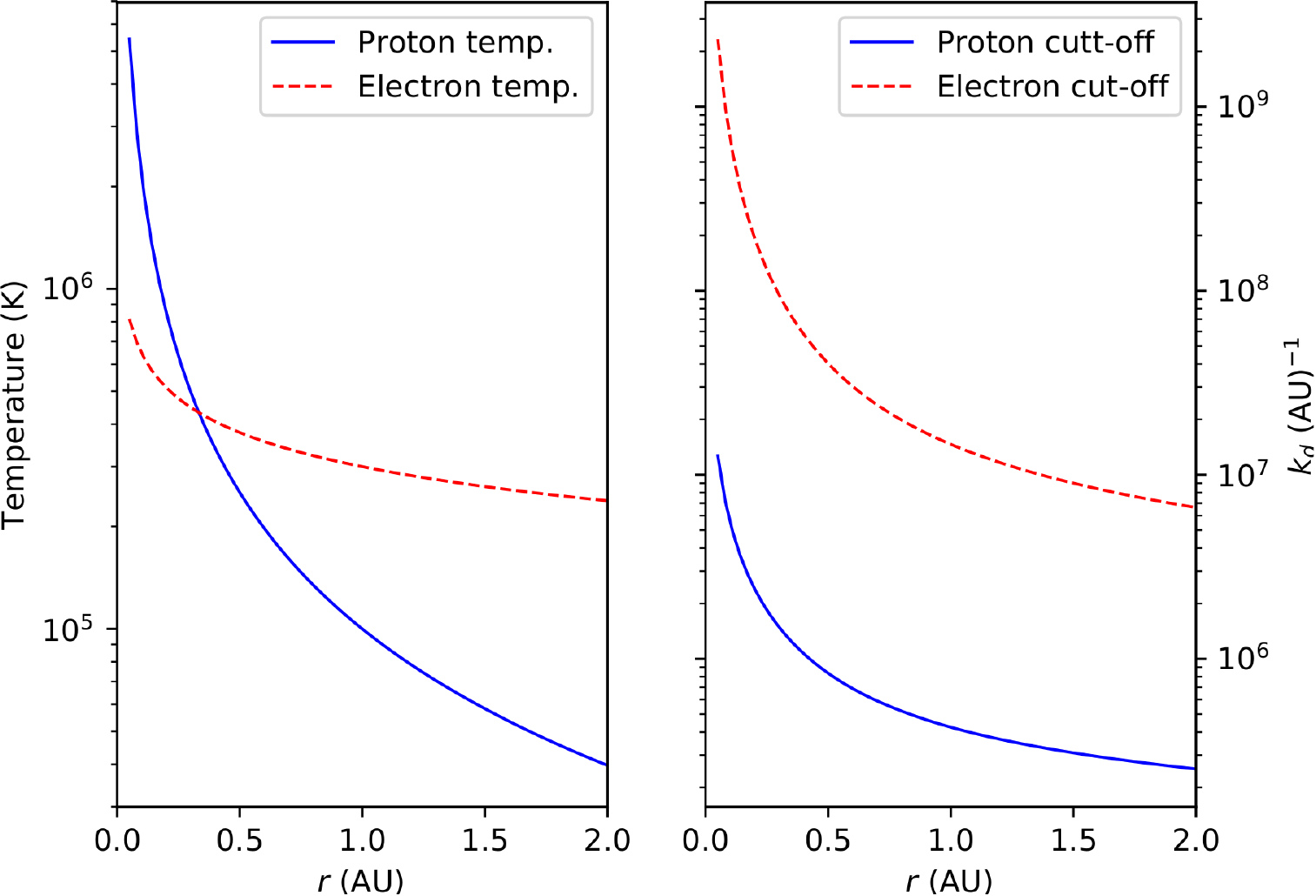}
\caption{The left panel shows the assumed radial dependence of the proton (solid blue) and electron (dashed red) solar wind plasma tempratures. The right panel shows the coresponding wave-number where turbulence dissipation is believed to start being effective. \label{fig:temps}}
\end{center}
\end{figure*}

\subsection{Nature of the assumed turbulence}

{Following \citet{shalchibook}, we assume the fluctuating field consists of a {\it slab} (with fluctuations directed along the mean field, $k_{||}$) and {\it 2D} (with fluctuations directed perpendicular to the mean field, $k_{\perp}$) component \citep[see also][]{Matthaeusetal1995}, so that the fluctuations can be described as

\begin{equation}
\vec{b}(x,y,z) = \vec{b}_{\mathrm{slab}}(z) + \vec{b}_{\mathrm{2D}}(x,y)
\end{equation}




For the total variance of the fluctuations, it follows that

\begin{equation}
\delta B^2 = \delta B_{\mathrm{slab}}^2 + \delta B_{\mathrm{2D}}^2.
\end{equation}

We work in terms of the so-called {\it wave approach}, where the slab fluctuations are considered to consist of a spectrum of (magneto-hydrodynamic) MHD waves with a wave frequency of $\omega$ and a growth/damping rate of $\gamma'$ \citep[so-called {\it plasma wave turbulence};][]{schlikkie}. In this model, we can simply specify any component of the turbulence, in wave-number space, as

\begin{equation}
\label{Eq:wave_correlation}
b_{\mathrm{slab},x}(k_{||},t) = b_{\mathrm{slab},x}(k_{||}) e^{-i \omega t + \gamma' t}.
\end{equation}

The strength of the background slab component is then calculated as

\begin{eqnarray}
\label{Eq:def_db_slab_aap}
\delta B_{\mathrm{slab}}^2 (t) &=& \langle  \vec{b}_{\mathrm{slab}}(k_{||},t) \cdot \vec{b}^*_{\mathrm{slab}}(k_{||},t) \rangle \\
&=& \langle  \vec{b}^2_{\mathrm{slab}}(k_{||}) \rangle e^{2 \gamma' t} \nonumber \\
&=& \int g^{\mathrm{slab}} (k_{||},t) \frac{\delta \left( k_{\perp} \right)}{k_{\perp}}  dk^3 \nonumber 
\end{eqnarray}

where $g(k_{||})$ is the so-called turbulence power spectrum and $\langle \cdot \rangle$ indicates an appropriate averaging procedure.

For this work, we consider only non-dispersive, parallel propagating and circularly polarized Alfv{\'e}n waves, where $\omega = V_A k_{||}$, and $V_A$ is the Alfv{\'e}n speed. This assumption is valid for near-relativistic electrons ($\sim 100$ keV) under consideration here: in Fig. \ref{fig:dispersion_diagram}, $k_{\mathrm{res}} \sim r_L^{-1}$ ($r_L$ being the maximal Larmor radius) indicates the maximum approximate wave-number at which these electrons will resonate, indicating that this is indeed in the Alfv\'en branch of each node.

It takes some care to interpret the growth/damping rate present in Eq. \ref{Eq:wave_correlation}. We assume this rate to consist of two contributions; the first is the growth/damping rate due to the streaming and non-thermal SEPs, given by $\gamma$, where, of course, $\gamma=0$ when there are no streaming particles. The dynamical character of the ``background turbulence" (that is, the quasi-stationary turbulence that is present in the undisturbed solar wind before the SEP event occurs), however, contributes to fluctuations changing at a dynamical timescale of $\tau$. Hence, we use

\begin{equation}
\gamma' = \tau^{-1} + \gamma.
\end{equation}

Following \citet{bieberetal1994}, we choose the dynamical timescale of the turbulence to be

\begin{equation}
\label{Eq:tau_define}
\tau^{-1} = \alpha \left| \omega \right| =  \alpha \left| V_A k_{||} \right|,
\end{equation}

where $\alpha = 1$ is chosen throughout for maximal dynamical effects. The SEPs are therefore introduced into a fully evolved turbulent medium in an equilibrium state. This leads to time-dependent changes in the level of the slab component to be solely due to wave growth/damping by non-thermal SEPs, i.e.

\begin{equation}
\label{Eq:define_growth_rate}
\frac{1}{g^{\mathrm{slab}} (k_{||},t)} \frac{\partial g^{\mathrm{slab}} (k_{||},t)}{\partial t} \approx 2 \gamma (k_{||},t).
\end{equation}

Assuming that the gyrotropic slab wave field includes forward ($j = +1$) and backward ($j = -1$) propagating waves with either a right-handed ($n = -1$) or left-handed polarization ($n = +1$), Eq. \ref{Eq:def_db_slab_aap} reduces to

\begin{equation}
\label{Eq:def_db_slab}
\delta B_{\mathrm{slab}}^2 = 2 \pi  \int_{0}^{\infty} \sum_{j = \pm 1} \sum_{n = \pm 1}  g^{\mathrm{slab}}_{n,j}(k_{||}) dk_{||}.
\end{equation}

}

The ratios of the different wave modes determine the cross helicity (ratio of forward to backward propagating waves),

\begin{equation}
\sigma_c \left(k_{||} \right) = \frac{\sum_{n = \pm 1}g^{\mathrm{slab}}_{n,j=+1} - \sum_{n = \pm 1}g^{\mathrm{slab}}_{n,j=-1} }{\sum_{n = \pm 1}g^{\mathrm{slab}}_{n,j=+1} + \sum_{n = \pm 1}g^{\mathrm{slab}}_{n,j=-1}},
\end{equation}

and the magnetic helicity (ratio of left hand to right hand polarized waves)

\begin{equation}
\sigma_m \left(k_{||} \right) =  \frac{\sum_{j = \pm 1}g^{\mathrm{slab}}_{n=+1,j}  - \sum_{j = \pm 1}g^{\mathrm{slab}}_{n=-1,j} }{\sum_{j = \pm 1}g^{\mathrm{slab}}_{n=+1,j} + \sum_{j = \pm 1}g^{\mathrm{slab}}_{n=-1,j}}.
\end{equation}

{For the 2D component we again include dynamical effect by assuming these are caused by perpendicular propagating Alfv{\'e}n waves, so that we may write

\begin{equation}
b_{\mathrm{2D},x}(k_{||},t) = b_{\mathrm{2D},x}(k_{||}) e^{-i (V_A k_{\perp}) t},
\end{equation}

where, similar to the slab component, the 2D fluctuations are in a dynamic, but quasi-steady state,

\begin{equation}
\frac{1}{g^{\mathrm{2D}} (k_{\perp},t)} \frac{\partial g^{\mathrm{2D}} (k_{\perp},t)}{\partial t} \approx 0,
\end{equation}

with the total strength thereof calculated as

\begin{eqnarray}
\delta B_{\mathrm{2D}}^2 &=& \int g^{\mathrm{2D}} (k_{\perp}) \frac{\delta \left( k_{||} \right)}{k_{\perp}}  dk^3 \nonumber \\
&=& 2\pi \int_{0}^{\infty} g^{\mathrm{2D}}(k_{\perp}) dk_{\perp}.
\end{eqnarray}

}







\begin{figure}
\begin{center}
\includegraphics[width=80mm]{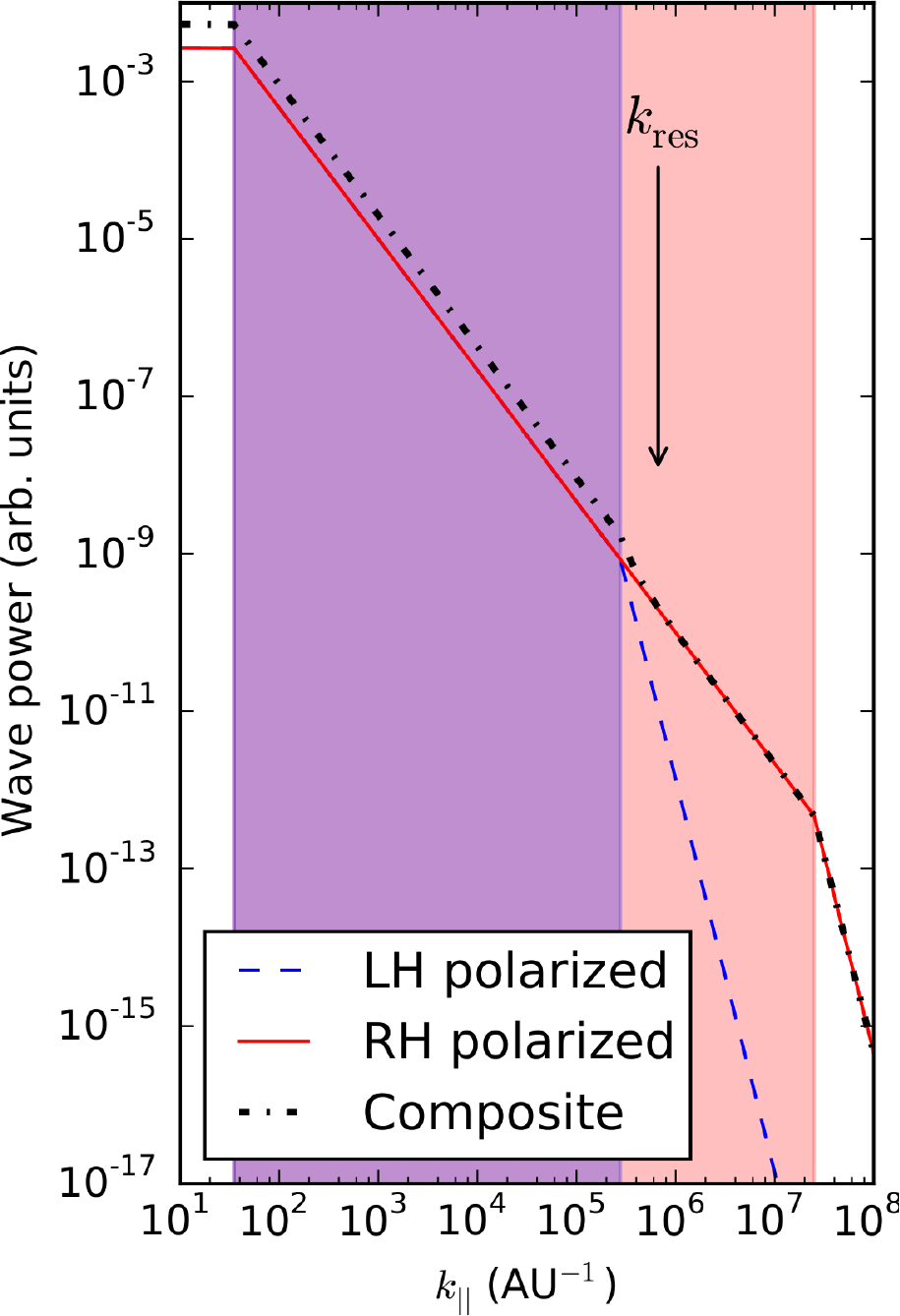}
\caption{The energy spectrum of the left-handed (dashed blue line) and right-handed (solid red line) slab waves. The total composite spectrum is shown by the dotted black line. Note that the onset of the dissipation range occurs for lower wave-numbers for the left-handed population. The approximate resonant scale, $k_{\mathrm{res}} \sim r_L^{-1}$, is shown for reference. \label{fig:turbulence_spectra}}
\end{center}
\end{figure}

\begin{figure*}
\begin{center}
\includegraphics[width=120mm]{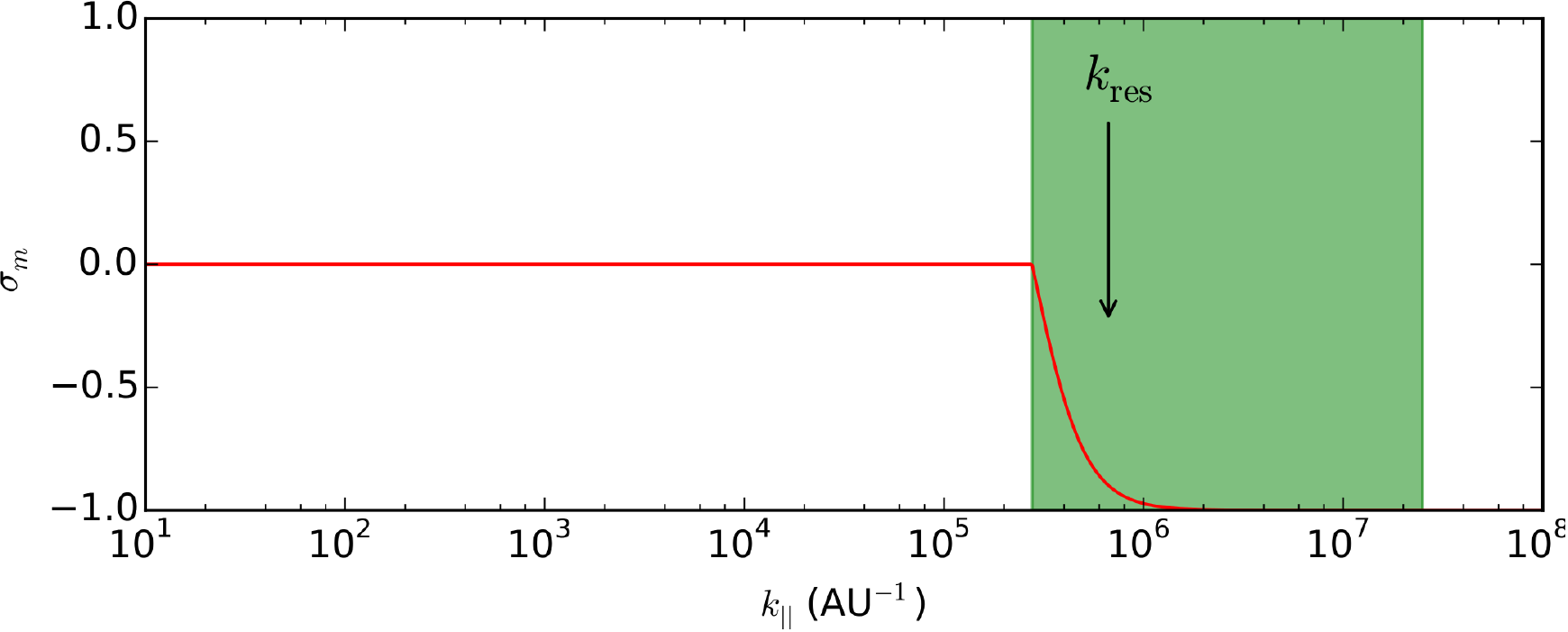}
\includegraphics[width=120mm]{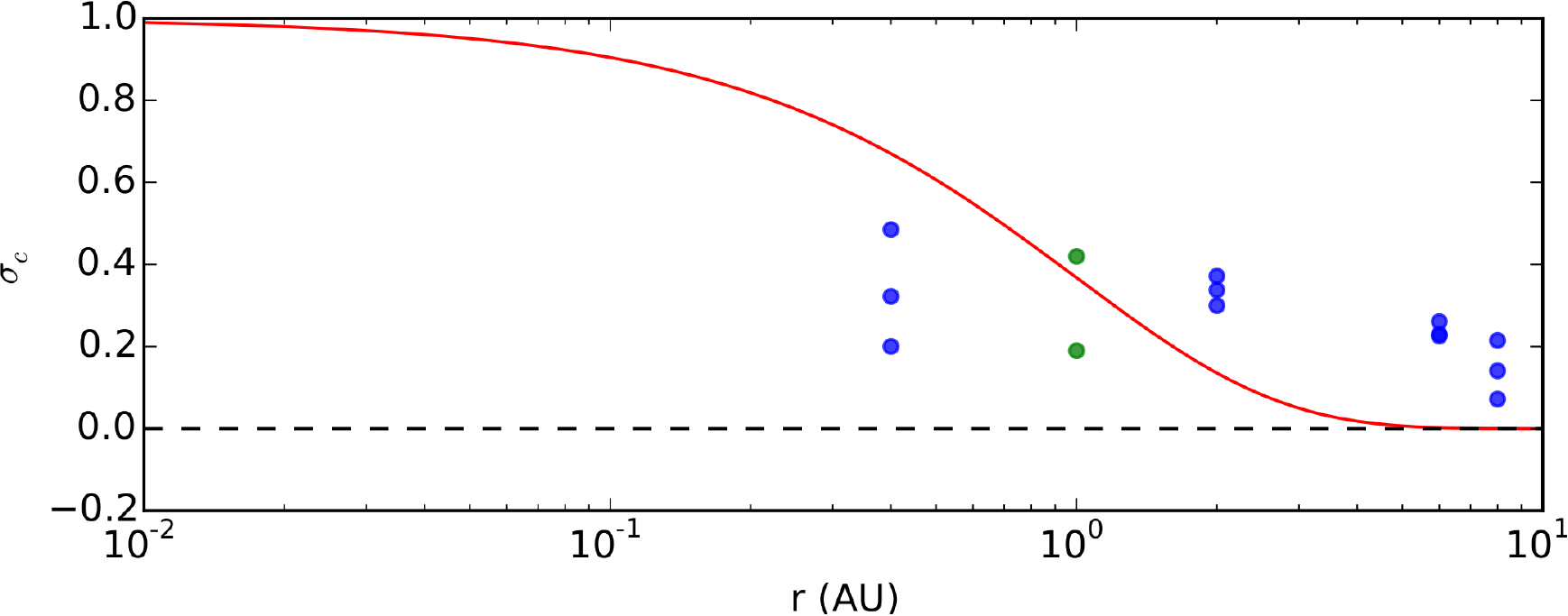}
\caption{The top panel shows the resulting magnetic helicity due to our assumptions of the different slab turbulence spectra. The approximate resonant scale, $k_{\mathrm{res}} \sim r_L^{-1}$, is again shown for reference. The bottom panel shows the assumed radial dependence of the cross-helicity. \label{fig:turbulence_helicities}}
\end{center}
\end{figure*}

\subsection{Onset and importance of the dissipation range}

In the inner heliosphere, low energy cosmic ray electrons resonate with the slab turbulence spectra near the so-called dissipation range \citep[][]{EB2015}. This is where the cascading waves start being heavily damped by e.g. a cyclotron resonance with the thermal plasma particles \citep[e.g.][]{woodhametal2018}. The assumed form of this part of the spectrum is therefore a crucial modelling component to correctly simulate the propagation of these electrons.

Our assumed slab turbulence spectrum is shown in Fig. \ref{fig:turbulence_spectra}. This form is exactly the same as used by \citet{straussetal2017}, but with a modified value of $k_d$, the wavenumber corresponding to the onset of the turbulence dissipation range, based on the model of \citet{ES2018}, given by

\begin{equation}
\label{Eq:k_d_onset}
k^{\mathrm{LH,RH}}_d \approx \frac{|\Omega|_{p,e}}{V_A + 3 v^{p,e}_{\mathrm{th}}},
\end{equation}

where we assume that left handed waves are damped by gyroresonant interactions with thermal protons and right handed waves interact with thermal electrons. The regions where this interaction is believed to be very effective, are contained within

\begin{equation}
\omega^{\mathrm{LH,RH}} \in |\Omega|_{p,e} \pm 3 k_{||} v^{p,e}_{\mathrm{th}} 
\end{equation}

for LH and RH waves separately \citep[see also the simulations by][]{cedricetal2017} and indicated on Fig. \ref{fig:dispersion_diagram} as the grey bands. The wavenumber where the grey bands intersect the non-dispersive wavemodes, are believed to be where wave damping becomes significant, and is given by Eq. \ref{Eq:k_d_onset}.

To use the expressions above, we also need an approximation for the electron and proton temperatures in the solar wind. These are assumed polytropic, 

\begin{equation}
T_{p,e} = T^0_{p,e} \left( \frac{r}{r_0} \right)^{-\chi_{p,e}}
\end{equation}

with $r_0 = 1$ AU and $ T^0_{p,e}$ a normalization value at Earth. Following \citet{cranmeretal2009}, we use $ T^0_{p}=30^5$ K and $ T^0_{e}=10^5$ K. The proton temperature is assumed to decrease adiabatically, $\chi_{p}=4/3$, while the electrons decrease is assumed to be more isothermal, $\chi_{e}=1/3$ \citep[e.g.][]{sittiescuddy1980,phillipsetal1995}. The resulting temperature profiles are shown in the left panel of Fig. \ref{fig:temps}.

Using these assumed temperature profiles, the resulting dissipation range onset are shown in the right panel of Fig. \ref{fig:temps} as a function of radial distance. Note that the dissipation onset that corresponds to LH polarized waves (and damped by protons) is always much smaller than the resulting quantity for RH polarized waves. This difference results in different forms of the slab turbulence spectrum; see Fig. \ref{fig:turbulence_spectra} where the LH-polarized inertial range (coloured blue in the figures) is much more narrow than the RH-polarized inertial range (the combined blue and red region). This will result in a non-zero magnetic helicity (shown in the top panel of Fig. \ref{fig:turbulence_helicities}) with electrons scattered, in this region, predominantly by RH polarized waves.

\subsection{Magnetic and cross helicities}

The magnetic helicity is calculated from the turbulence spectra discussed in the previous paragraph and shown in the top panel of Fig.\ref{fig:turbulence_helicities}. At high wave-numbers this value becomes non-zero and indicates, for our assumptions, an excess of RH polarized slab waves.

We model the background cross-helicity by a simple exponential function,

\begin{equation}
\sigma_c = \exp \left[- \frac{r}{r_0} \right],
\end{equation}

which is independent of wavenumber and with $r_0=1$ AU. This assumed form is shown in the bottom panel of Fig. \ref{fig:turbulence_helicities} along with the observations of \citet{robertsetal1987a,robertsetal1987b} (blue dots) and \citet{breechetal2005} (green dots). This form is motivated by the stimulations of e.g. \citet{adhikari} that shows how the highly anisotropic slab wave-field at the Alfv\'en radius (only forward moving Afv{\'e}n waves can escape this surface and be convected by the solar wind into the inner heliosphere) can become increasingly isotropic near Earth's radius due to e.g. wave-wave interactions.

Although seemingly insignificant, the effects of a non-zero magnetic and cross-helicity has important implication for particle transport as discussed later on. The upcoming missions, {\it Parker Solar Probe} and {\it Solar Orbiter}, will be able to measure these quantities in detail.

\section{Numerical transport model}

The propagation of energetic electrons is described by the so-called focussed transport equation \citep[e.g.][]{skilling1971}, given by 

\begin{eqnarray}
\label{Eq:TPE}
\frac{\partial f}{\partial t} &=& - \nabla \cdot \left( \mu v \hat{b} f \right) - \frac{\partial}{\partial \mu} \left( \frac{1-\mu^2}{2L} vf \right) 
 \nonumber \\  
&+& \frac{\partial}{\partial \mu} \left(D_{\mu\mu}  \frac{\partial f}{\partial \mu} \right) +  \nabla \cdot \left( \mathbf{D}^{(x)}_{\perp}\cdot \nabla f \right) 
\end{eqnarray}

and solved by means of the numerical approach outlined by \citet{straussfichtner2015} to yield the gyro-tropic particle distribution function $f$. In Eq. \ref{Eq:TPE}, $\hat{b}$ is a unit vector pointing along the mean heliospheric magnetic field, $v$ is the particle speed, $\mu$ is the cosine of the pitch-angle, $ \mathbf{D}^{(x)}_{\perp}$ contains the perpendicular diffusion coefficients and is specified in spherical coordinates (radial distance, $r$, and azimuthal angle, $\phi$), $D_{\mu \mu}$ is the pitch-angle diffusion coefficient and the focusing length is calculated as 

\begin{equation}
L^{-1} = \nabla \cdot \hat{b},
\end{equation}

for a \citet{parker} heliospheric magnetic field (HMF), normalized to 5 nT at Earth. The first order anisotropy is calculated as

\begin{equation}
A(r,\phi,t) = 3 \frac{\int_{-1}^{+1}  \mu f  d\mu}{\int_{-1}^{+1} f   d\mu},
\end{equation}

and the omni-directional distribution function as

\begin{equation}
F(r,\phi,t) =\frac{1}{2} \int_{-1}^{+1}   f(r,\phi,\mu,t)  d\mu.
\end{equation}

From $F$, the isotropic differential intensity is calculated as $j = p^2 F$, with $p$ being particle momentum.

{As an inner boundary condition, the following mono-energetic isotropic injection function

\begin{eqnarray}
f(r&=&r_0,\phi,t, E) = C \frac{E_0 \delta \left( E - E_0 \right)}{t}   \nonumber \\
  & \times &  \exp \left[ - \frac{(\phi - \phi_0)^2}{2 \sigma^2}  \right]   \exp \left[ -\frac{\tau_a}{t}  -\frac{t}{\tau_e}  \right] \label{Eq:reid_injection}
  \end{eqnarray}

is prescribed at the inner boundary, located at $r_0 = 0.05$ AU. Gaussian injection in $\phi$ is assumed with $\phi_0 = \pi/2$ and $\sigma$ determining the broadness thereof. The model is solved for $E_0 = 100$ keV.} We use a value of $\sigma = 20^{\circ}$ for the simulations presented here. It is believed that magnetic structures near the Sun \citep[e.g.][]{kleinetal2008} can significantly broaden the compact source regions from a solar flare below the Alfv{\'e}nic radius. We can account for this broadening in our model by specifying $\sigma$, although the simulations of \citet{straussetal2017} have shown that the modelled intensities are relatively insensitive to the choice of this parameter as long as significant levels of perpendicular diffusion is present. A Reid-Axford \citep[][]{reid1964} temporal injection profile is specified with $\tau_a = 1/10$ hr, $\tau_e = 1$ hr, and $C$ is a parameter that is adjusted to give the correct intensity at Earth. Other important quantities assumed in the model are: a solar wind number density of 5 particles.cm$^{-3}$ at Earth decreasing as $r^{-2}$ and a constant solar wind speed of $V_{sw}=400$ km.s$^{-1}$.

The pitch-angle and perpendicular diffusion coefficients, needed as input to Eq. \ref{Eq:TPE}, are discussed and calculated in the next section. 

\section{Transport coefficients}

\begin{figure*}
\begin{center}
\includegraphics[width=120mm]{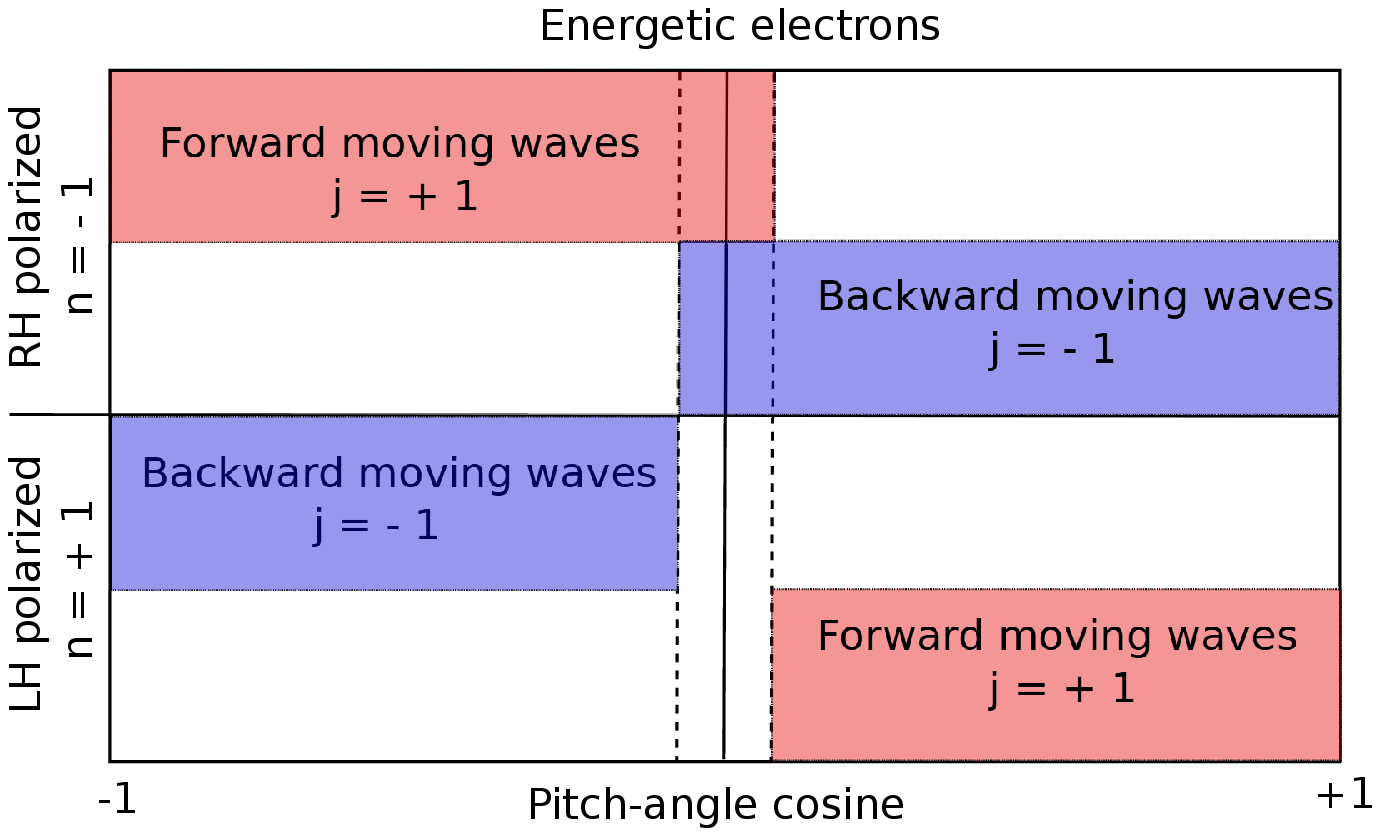}
\caption{The figures show when it is possible for SEP electrons to resonante with either wave population. Shaded areas indicate regions of possible resonance, while the dashed region indicates the classical resonance gap that is present when either only LH polarized waves are considered, or if the waves are considered stationary. \label{fig:when_resonante}}
\end{center}
\end{figure*}

\begin{figure*}
\begin{center}
\includegraphics[width=160mm]{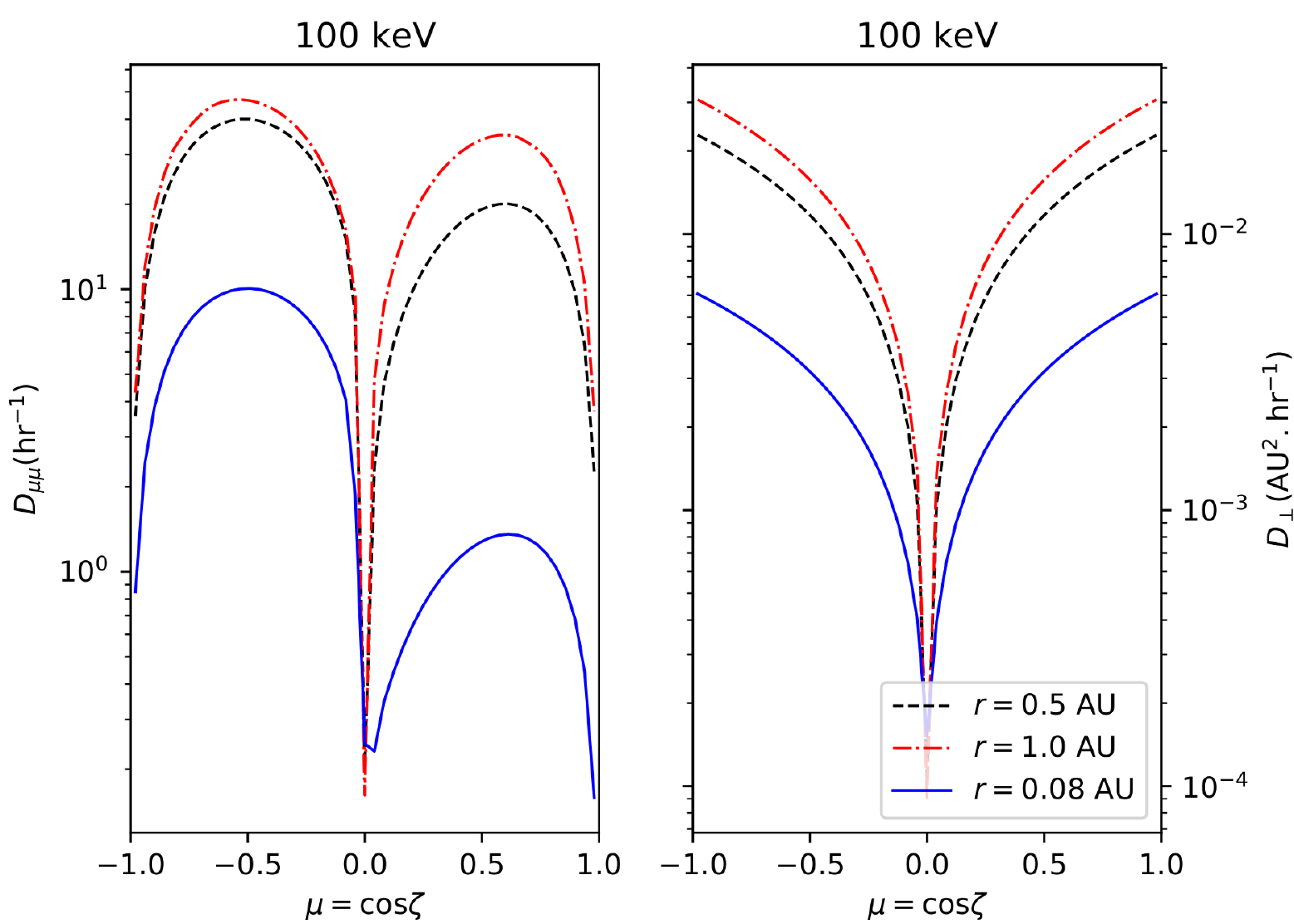}
\caption{The left panel shows the calculated $D_{\mu \mu}$, at different radial positions, as a function of pitch-angle. The right panel shows the equivalent form of $D_{\perp}$.  \label{fig:duu}}
\end{center}
\end{figure*}

\begin{figure}
\begin{center}
\includegraphics[width=80mm]{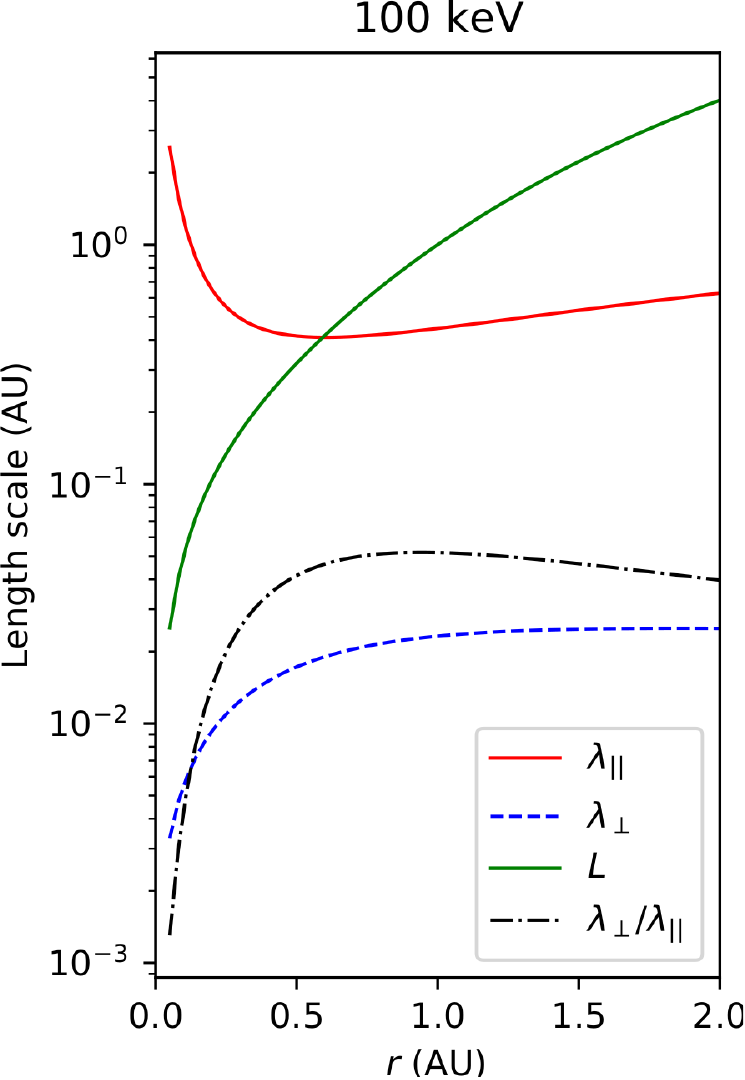}
\caption{The resulting $\lambda_{||}$ (solid red line) and $\lambda_{\perp}$ (dashed blue line) as a function of radial distance. Also shown is the radial dependence of the focusing lenght (solid green line) and the ratio $\lambda_{\perp}/\lambda{||}$ (dash-dotted balck line). \label{fig:lambdas}}
\end{center}
\end{figure}

In this work we try to to use, for as far as possible, a ``physics-first" approach to simulation SEP transport. That is, we attempt to specify, as self-consistently as possible, the background turbulence spectra, and calculate the SEP transport coefficients from these assumed forms.
\subsection{Pitch-angle diffusion}

The plasma wave pitch-angle diffusion coefficient is given by e.g. \citet{schlikkie} as

\begin{equation}
D_{\mu \mu} = \frac{\pi v^2 (1 - \mu^2)}{B_0^2 r_L^2} \int_0^{\infty} dk_{||} \sum_{n = \pm 1} \sum_{j=\pm1} \mathcal{R}_{n,j} g^{\mathrm{slab}}_{n,j},
\end{equation}

where we have assumed that $V_A \ll v$. The resonance functions are given by

\begin{equation}
\mathcal{R}_{n,j} = \frac{ \tau^{-1} }{\left( \tau^{-1} \right)^2 + \left( v \mu j k_{||} - \omega + n\Omega \right)^2},
\end{equation}

with $\gamma \ll \tau^{-1}$; the broadening of the resonance function is therefore assumed to be entirely due to the dynamic character of the background fluctuations. This is the form of $D_{\mu \mu}$ that will be used for the rest of the study.

We can use this expression to see when electrons will resonate with what type of slab wave (i.e. when is $\mathcal{R}_{n,j} > 0$). This is shown in Fig. \ref{fig:when_resonante}, where the shaded regions indicate possible resonance with either forward (red regions) or backward (blue regions) propagating waves with either polarization state. The vertical dashed lines show the classical resonance gap which is present when either only LH-polarized waves are assumed, or if the waves are assumed static. Note that when a beam of electrons is present (i.e. $\mu > 0)$, these particles will strongly resonate with forward propagating LH waves and backward propagating RH polarized waves.

The resulting $D_{\mu \mu}$ is shown in the left panel of Fig. \ref{fig:duu} at three different radial positions. Note that this coefficient is not symmetric about $\mu \sim 0$, but enhanced for backwards moving particles. This is a direct result of the non vanishing magnetic and cross-helicities assumed in the model; there is an excess of RH-polarized, forward propagating waves in the model, as therefore backwards propagating particles experience more scattering. Refer again to Fig. \ref{fig:when_resonante}. Near Earth, $D_{\mu \mu}$ becomes more symmetrical as the cross-helicity decreases to zero.

The parallel mean free path (MFP, $\lambda_{||}$) is calculated from $D_{\mu \mu}$ following the usual definition of \citet{hasselmaanwibberenz1968},

\begin{equation}
\lambda_{||} = \frac{3v}{8} \int_{-1}^{+1} \frac{\left( 1 - \mu^2 \right)^2}{D_{\mu\mu}} d\mu .
\end{equation}

The resulting $\lambda_{||}$ is shown in Fig. \ref{fig:lambdas} as a function of radial distance. {Note the strong radial dependence of $\lambda_{||}$ and the large values thereof close to the Sun.}

If we, however, consider the case of negligible damping (ND) or dynamical effects, the resonance function reduces to

\begin{equation}
\label{Eq:non_damped_1}
\mathcal{R}^{\mathrm{ND}}_{n,j} \approx \pi \delta \left( v \mu j k_{||} - \omega + n\Omega \right),
\end{equation}

and the pitch-angle diffusion coefficient becomes

\begin{equation}
D^{\mathrm{ND}}_{\mu \mu} = \frac{\pi^2 v (1 - \mu^2)}{B_0^2 r_L^2 |\mu|}  \sum_{n = \pm 1} \sum_{j=\pm1} g_{slab}^{n,j}(k_{||}^{\mathrm{res}}) ,
\end{equation}

where we have introduced the resonant wavenumber as

\begin{equation}
\label{Eq:non_damped_3}
k_{||}^{\mathrm{res}} = \frac{n\Omega}{V_A -  v\mu j }.
\end{equation}

This approximate form of the pitch-angle diffusion coefficient, $D^{\mathrm{ND}}_{\mu \mu}$, will be used later on to derive the wave damping/growth rate.

\subsection{Perpendicular diffusion}

No consistent theory exists to describe perpendicular diffusion, on the pitch-angle level, in the presence of dynamical turbulence. However, we may include the dynamic contribution of propagating Alfv{\'e}n waves (or rather, particles drifting in the random electric fields induced by the propagating Alfv{\'e}n waves) by using the approximate result of \citet{straussetal2016}

\begin{equation}
\label{Eq:approx_flrw}
D_{\perp}^{\mathrm{dyn}} \approx  \langle l_{\perp} \rangle \left(v |\mu| + V_A \right) \frac{\left( \delta B^2_{\mathrm{2D}} \right)}{B_0}^{1/2} ,
\end{equation}

which is valid for the limit of $r_L \ll \langle l_{\perp} \rangle$, where $\langle l_{\perp} \rangle$ is, approximately, the perpendicular correlation scale. In the magnetostatic limit of $V_A = 0$, Eq. \ref{Eq:approx_flrw} should, however, converge to the field-line random walk (FLRW) model of \citet{jokipii1966}, where

\begin{equation}
D_{\perp}^{\mathrm{FLRW}} = a v |\mu| \kappa_{\mathrm{FL}},
\end{equation}

with $a^2 \in \left[ 0, 1 \right]$ a parameter determining the probability of particles being stuck to wandering field-lines, and the field-line diffusion coefficient $\kappa_{\mathrm{FL}}$ given by e.g. \citet{qinshalchi2014} as

\begin{equation}
\kappa^2_{\mathrm{FL}} = \frac{\pi}{B_0^2} \int_0^{\infty} k^{-2}_{\perp} g^{\mathrm{2D}} (k_{\perp})  dk_{\perp}.
\end{equation}

Making the rather {\it ad hoc} identification, 

\begin{equation}
\kappa_{\mathrm{FL}} \stackrel{?}{=}  \langle l_{\perp} \rangle  \frac{\left( \delta B^2_{\mathrm{2D}} \right)}{B_0}^{1/2}, 
\label{Eq:aap_adhoc}
\end{equation}

which is remarkably similar to the form used by \citet{2007_Mattheaus_etal_ApJ}, but given, by these authors, in terms of the ultrascale, we use the following estimate for the dynamical perpendicular diffusion coefficient,

\begin{equation}
D_{\perp}^{\mathrm{dyn}} \approx  \left( a v |\mu| + \alpha V_A \right) \kappa_{\mathrm{FL}}.
\end{equation}

The value of $a$ is changed in later sections to illustrate the role of perpendicular diffusion in wave amplification, while $\alpha$ was introduced for consistency with Eq. \ref{Eq:tau_define}.\\

The isotropic perpendicular mean free path, is calculated as

\begin{equation}
\lambda_{\perp} = \frac{3}{2v} \int_{-1}^{+1} D_{\perp} d \mu.
\end{equation}

Both $D_{\perp}$ and $\lambda_{\perp}$ are shown in Figs. \ref{fig:duu} and \ref{fig:lambdas}. Note that the inclusion of dynamical effects, although small, leaves to a non-zero value of $D_{\perp}$ near $\mu \sim 0$. This effect may, however, become increasingly dominant at lower energies where $v \sim V_A$.

It should also be noted that if $g^{\mathrm{2D}}(k_{\perp})$ is interpreted as the probability of finding a fluctuation at a given wave-number with a given energy (amplitude), then the characteristic length scale (which can be identified as the correlation length) can be evaluated as a root-mean-squared average scale using the second-order moment of the distribution

\begin{equation}
\langle l_{\perp} \rangle ^2  \sim \langle k^{-2}_{\perp} \rangle = \frac{\int k^{-2}_{\perp} g^{\mathrm{2D}} (k_{\perp}) dk_{\perp}}{\int  g^{\mathrm{2D}} (k_{\perp}) dk_{\perp}}
\end{equation}

which leads to Eq. \ref{Eq:aap_adhoc} after some re-arrangement.\\

\section{Wave growth/damping and its numerical implementation}

\begin{figure*}
\begin{center}
\includegraphics[width=160mm]{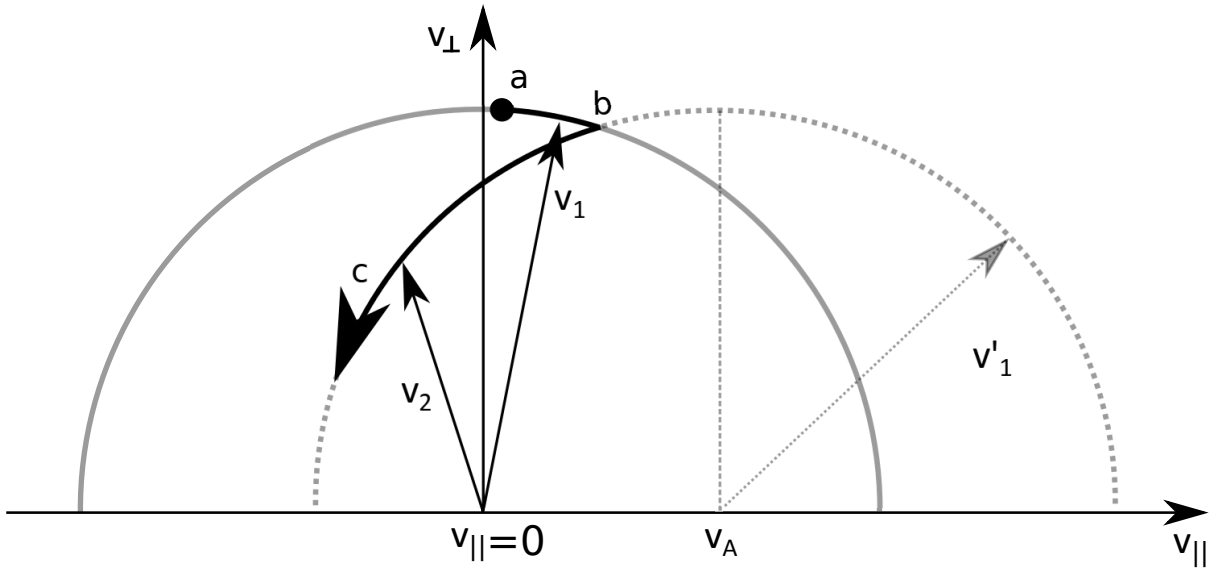}
\caption{An illustration of how pitch-angle scattering by forward propagating Alfv{\'e}n waves can lead to wave growth. See the text for details. \label{fig:wave_growth}}
\end{center}
\end{figure*}

The basic idea behind wave growth/damping is illustrated in Fig. \ref{fig:wave_growth}: a distribution of particles with speed $v_1$ is isotropic with respect to the bulk flow frame (or, fixed frame) denoted by $v_{||}=0$ in the figure, i.e. their pitch-angles are equally populated so that they form a shell distribution (the grey half-circle on the figure). Forward propagating Alfv{\'e}n waves (with a speed of $v_{||} = V_A$) are now introduced into the system and they interact with the particles through gyro-resonance which leads to pitch-angle scattering. The Alfv{\'e}n waves will now attempt to isotropize the distribution with respect to the wave frame (the dashed grey half-circle on the figure). Note that the energy of particles, as observed in the wave frame, is conserved during the interaction, $v'_1 = v_1$. However, the particle energy, as measured in the flow frame, decreased to $v_2 < v_1$ as the particle moved from $a \rightarrow c$, with the scattering event at $b$ (although pitch-angle scattering in this sense will be a continuous process). Because of the conservation of energy, the energy that the particle lost was gained by the Alfv{\'e}n waves responsible for that scattering event, and hence we may state that forward propagating particles ($v_{||}>0$) that interact with forward moving Alfv{\'e}n waves ($v_{||} = v_A>0$) will loose energy during gyro-resonant interactions, and will in turn grow and/or amplify forward moving Alfv{\'e}n waves. A similar sketch and argument can be found in \citet{afani}.

Of course, the opposite interaction is also possible: backward propagating particles that interact with forward moving Alfv{\'e}n waves will gain energy during gyro-resonant interactions, and will in turn damp the forward moving Alfv{\'e}n waves. If the particle distribution is isotropic (equal number of forward and backward propagating particles), the net effect will be that the interactions average out, and no net energy change (for either the particles or the waves) will take place.

A more rigorous treatment of the wave growth/damping problem is given in Appendix \ref{Sec:appendix_1} and leads to the result

\begin{equation} 
\label{Eq:final_wavre_growth_results_agaion}
\gamma_{n,j} =  j \frac{\pi^2}{4} \frac{p_0^4}{k^{\mathrm{res}}_{||}} \frac{\Omega^2}{  \mathcal{B}} \frac{V_A}{v}   (1 - \mu_{\mathrm{res}}^2)  \left. \frac{\partial f }{\partial \mu} \right|_{\mu_{\mathrm{res}}} , 
\end{equation}

in terms of the resonant wavenumber ($k_{||}^{\mathrm{res}}$) and pitch angle ($\mu_{\mathrm{res}}$), and the energy contained in the background magnetic field, $\mathcal{B}$. By looking at this result, it is clear that for a isotropic distribution, $\partial f / \partial \mu \approx 0$, we should have $\gamma \approx 0$ and no net wave amplification will occur. Moreover, we should note that the growth rate is generally expected to be low, due to the presence of the factor $V_A / v \ll 1$. But, as we will show, wave growth may still reach appreciable levels if enough streaming particles are present, and if the amount of scattering is sufficient (this is quantified by the plasma quantities present in Eq. \ref{Eq:final_wavre_growth_results_agaion}), as the case $D_{\mu \mu} \approx 0$ will lead to $\gamma \approx 0$ (the particles cannot interact with the waves; see e.g. Eq. \ref{Eq:wave_growth_1}). Lastly, we note that the sign of $\gamma$ is determined by the wave propagation direction, $j$ in accordance with the sketch presented in Fig. \ref{fig:wave_growth}.

In the numerical model, at each timestep, the numerical derivative $\partial f/\partial \mu$ is calculated, which allows us to calculate the growth/damping rate as given by Eq. \ref{Eq:final_wavre_growth_results_agaion}. The different wave amplitudes are then adjusted accordingly by calculating

\begin{equation}
g^{\mathrm{slab}}_{n,j}(k^{\mathrm{res}}_{||},t) = g^{\mathrm{slab}}_{n,j}(k^{\mathrm{res}}_{||},t_0) \exp \left(2 \int_{t_0}^t   \gamma_{n,j}(k^{\mathrm{res}}_{||},t') dt' \right),
\end{equation}

or, approximated in its discretized form as

\begin{equation}
g^{\mathrm{slab}}_{n,j}(k^{\mathrm{res}}_{||},t_0+\Delta t) = g^{\mathrm{slab}}_{n,j}(k^{\mathrm{res}}_{||},t_0) \exp \left(2 \gamma_{n,j}(k^{\mathrm{res}}_{||},t_0) \cdot \Delta t \right),
\end{equation}

where the numerical timestep is $\Delta t = t - t_0$. {The time-dependent turbulent levels are then used to calculate time-dependent transport coefficients, which are then incorporated in the time-dependent simulations. A self-consistent description is therefore obtained.} {For the present study, transport effects of the newly generated waves, in both physical and wave-number space, are neglected. This is partly motivated by the fact that the SEP electrons propagate relativistically from the Sun to Earth and that the distribution peaks within minutes, while it is unlikely that any newly generated waves will propagate appreciably far within this timeframe. Although wave attenuation might therefore be important later in an SEP event, and can considerably change the characteristic of the generated turbulence, the effects thereof on SEP propagation can most likely be safely neglected. Wave cascading effects, or more importantly for electrons, wave dissipation effects, were assumed negligible following a similar argument; while it is unclear if these effects will be significant in the assumed timeframes, we also do not currently have a firm grasp of what the wave dissipation rate should be. Although outside the scope of the present study such effect will be included in future work.}

What do we expect for a beam (i.e. an excess of forward propagating particles, i.e. $\partial f/\partial \mu > 0$) of electrons? We expect to amplify forward moving waves and damp backward moving waves of both polarities. The wavenumber where this occurs (i.e. the resonant wavenumber) however will also depend on the polarity of the waves. Moreover, because forward moving particles resonate more with forward moving LH-polarized waves, we expect these waves to be preferably amplified.  Refer again to Fig. \ref{fig:when_resonante}.

\section{Results}

\subsection{Reference solution}

\begin{figure*}
\begin{center}
\includegraphics[width=140mm]{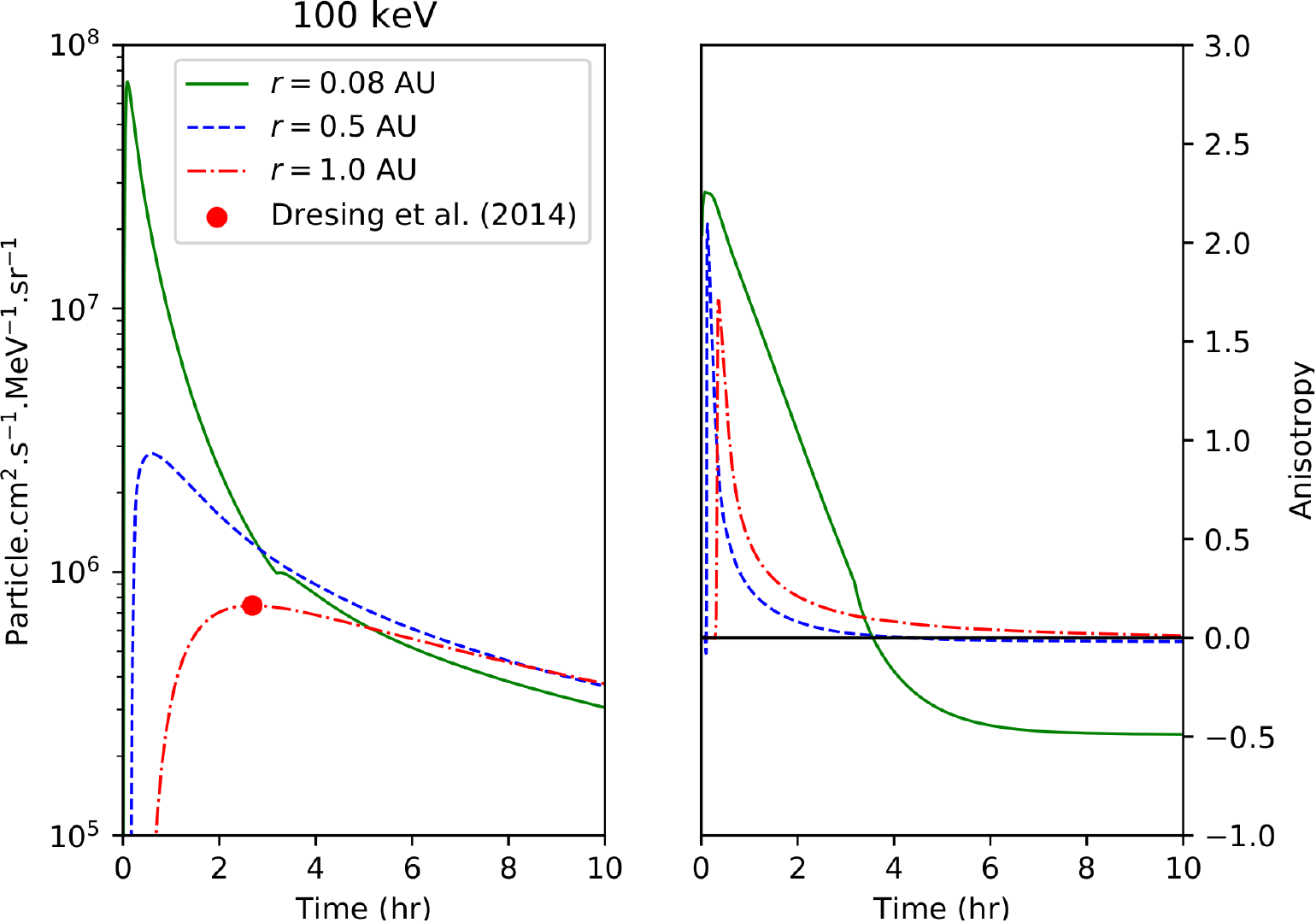}
\caption{The left panel shows the resulting temporal profile of the differential intensity at three different radial positions along the same magnetic field-line that connects to the SEP source. The position of these virtual spacecraft are indicated on Fig. \ref{fig:intensity_contour}. The maximum intensity at Earth is normalized to the maximum value reported by \citet{dresing2014} as indicated by the red circle. The right panel shows the corresponding anisotropies.   \label{fig:omni_time}}
\end{center}
\end{figure*}

\begin{figure*}
\begin{center}
\includegraphics[width=140mm]{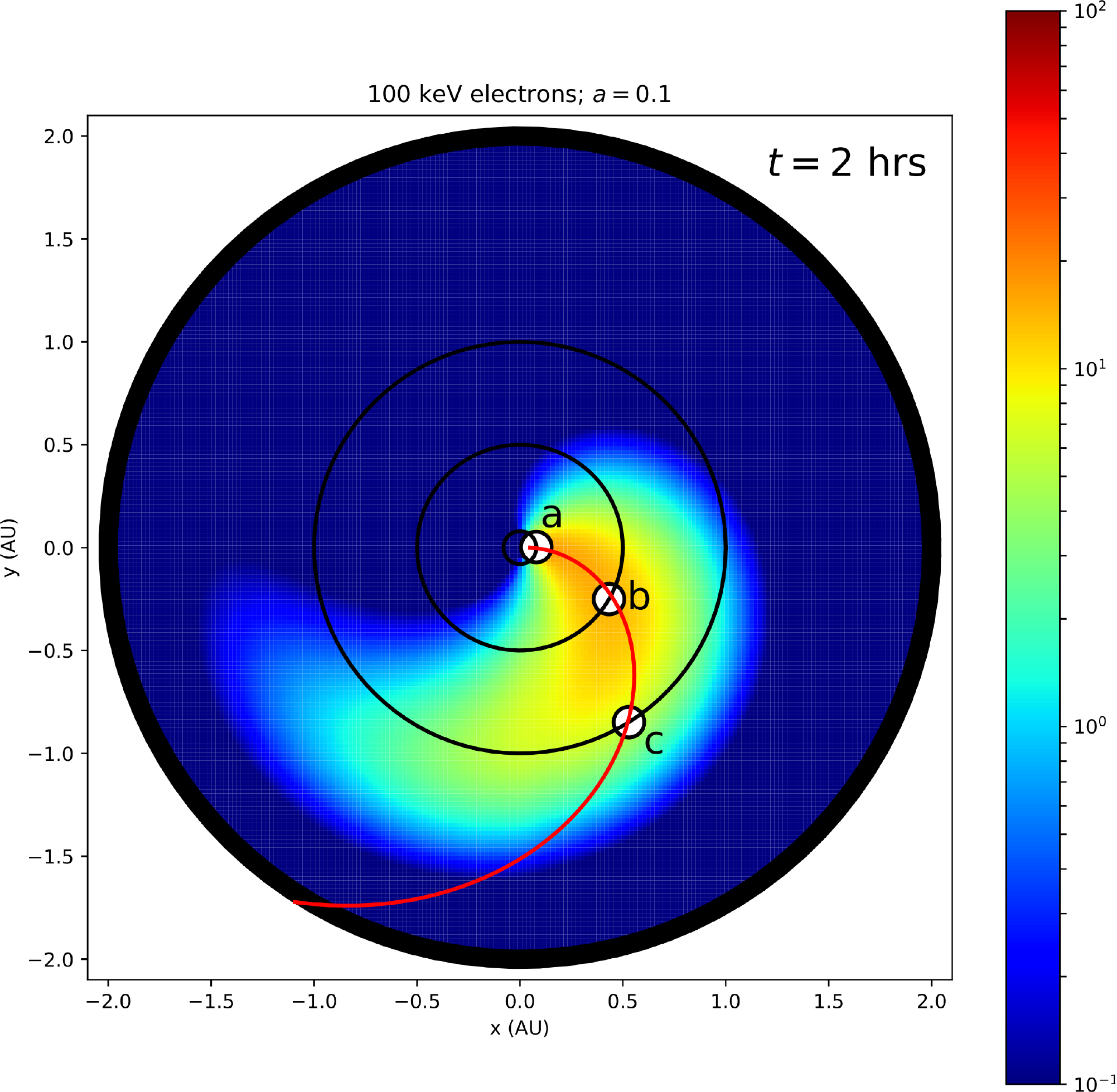}
\caption{The resulting omni-directional intensity (normalized to 100\% at its maximum), shown as a contour plot in the ecliptic plane, at $t=2$ hrs after the modelled SEP event. The position of the three virtual spacecraft used throughout are indicated, along with the magnetic fieldline that connects with to the maximum of the SEP source. \label{fig:intensity_contour}}
\end{center}
\end{figure*}

First, we show so-called reference solutions. These are the resulting model intensities without including any wave-growth in the model. It is extremely important to normalize these intensities to the correct levels, as the SEP intensity will directly influence the amount of wave modification. We start by injecting an arbitrary intensity of SEPs into the model, solve the model to obtain the intensity at Earth, and normalize this to observed levels. Note that the normalization is performed only once in order for the results to remain consistent. The temporal profile for this reference solution is shown in the left panel of Fig. \ref{fig:omni_time} for various radial positions; all magnetically connected to the SEP source, i.e. along the same magnetic field line. The maximum intensity at 1 AU is normalized to the maximum value reported by \citet{dresing2014} for electrons of these energies  (indicated on the figures by the red circle). The right panel of the figure shows the corresponding anisotropies. Interesting, the anisotropy near the Sun becomes negative after the initial injection phase. This is due to the large $\lambda_{||}$ value close to the Sun that does not effectively isotropizes the distribution; after the initial injection, SEPs are scattered and isotropized further outwards, where after they propagate back towards the Sun.

The differential intensity of the reference solution shown in Fig. \ref{fig:intensity_contour}, as a contour plot in the ecliptic plane, at $t=2$ hrs after the SEP events was injected into the model. The position of the three ``virtual spacecraft", where the intensities are calculated throughout this work and in Fig. \ref{fig:omni_time}, are also shown. Note that the azimuthal extent of this distribution will depend heavily on the effectiveness of perpendicular diffusion. The effect thereof is discussed in a later section.

\subsection{Wave growth/damping effects}

\begin{figure*}
\begin{center}
\includegraphics[width=160mm]{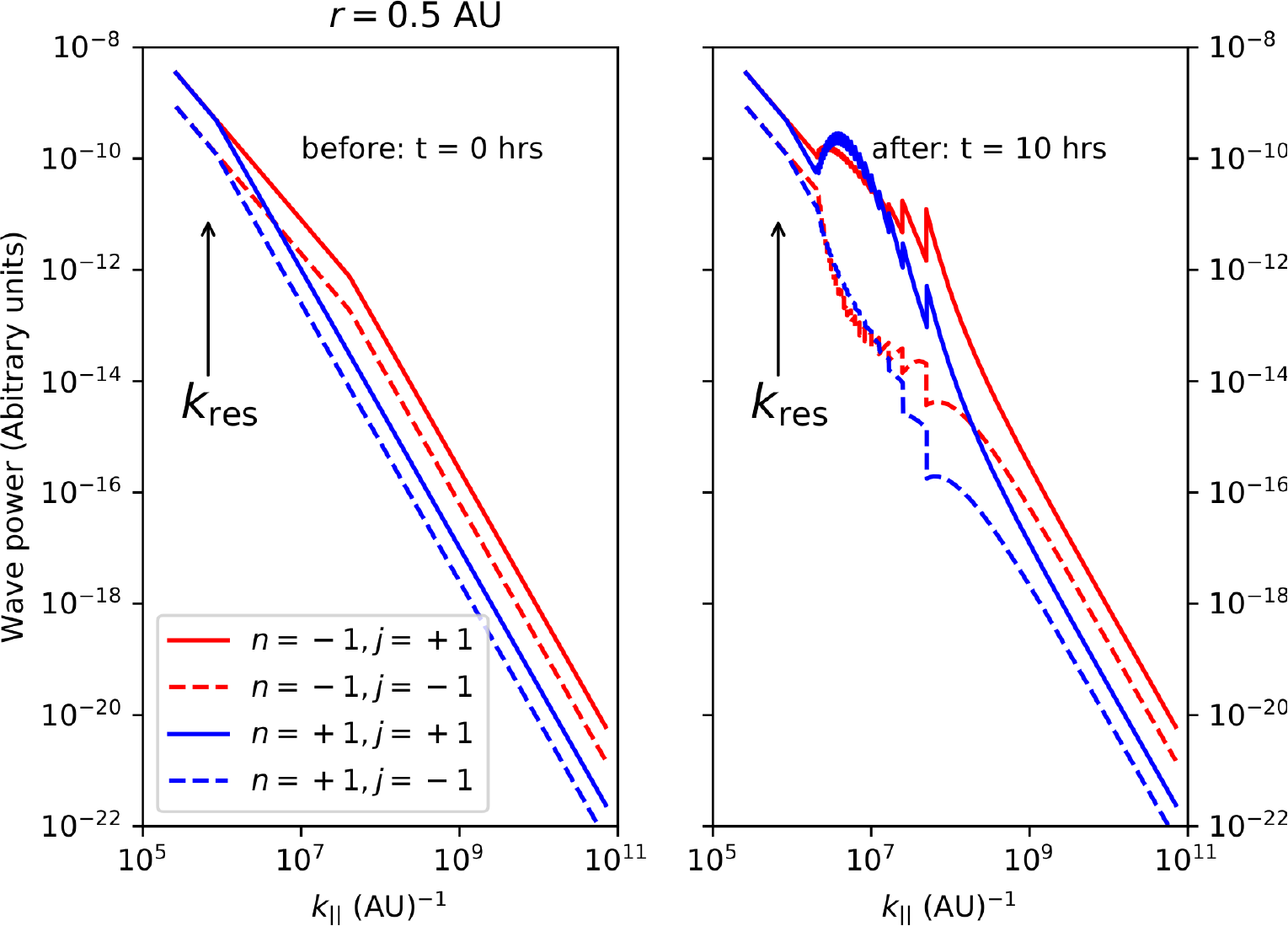}
\caption{The left panel shows the four components of the slab turbulence spectrum (at $r=0.5$ AU) before the SEP events was introduced into the model. The right panel shows the resulting spectrum, at $t=10$ hrs, after the SEP event sweaped past the virtual spacecraft. \label{fig:turb_spectra_after_event}}
\end{center}
\end{figure*}

\begin{figure*}
\begin{center}
\includegraphics[width=85mm]{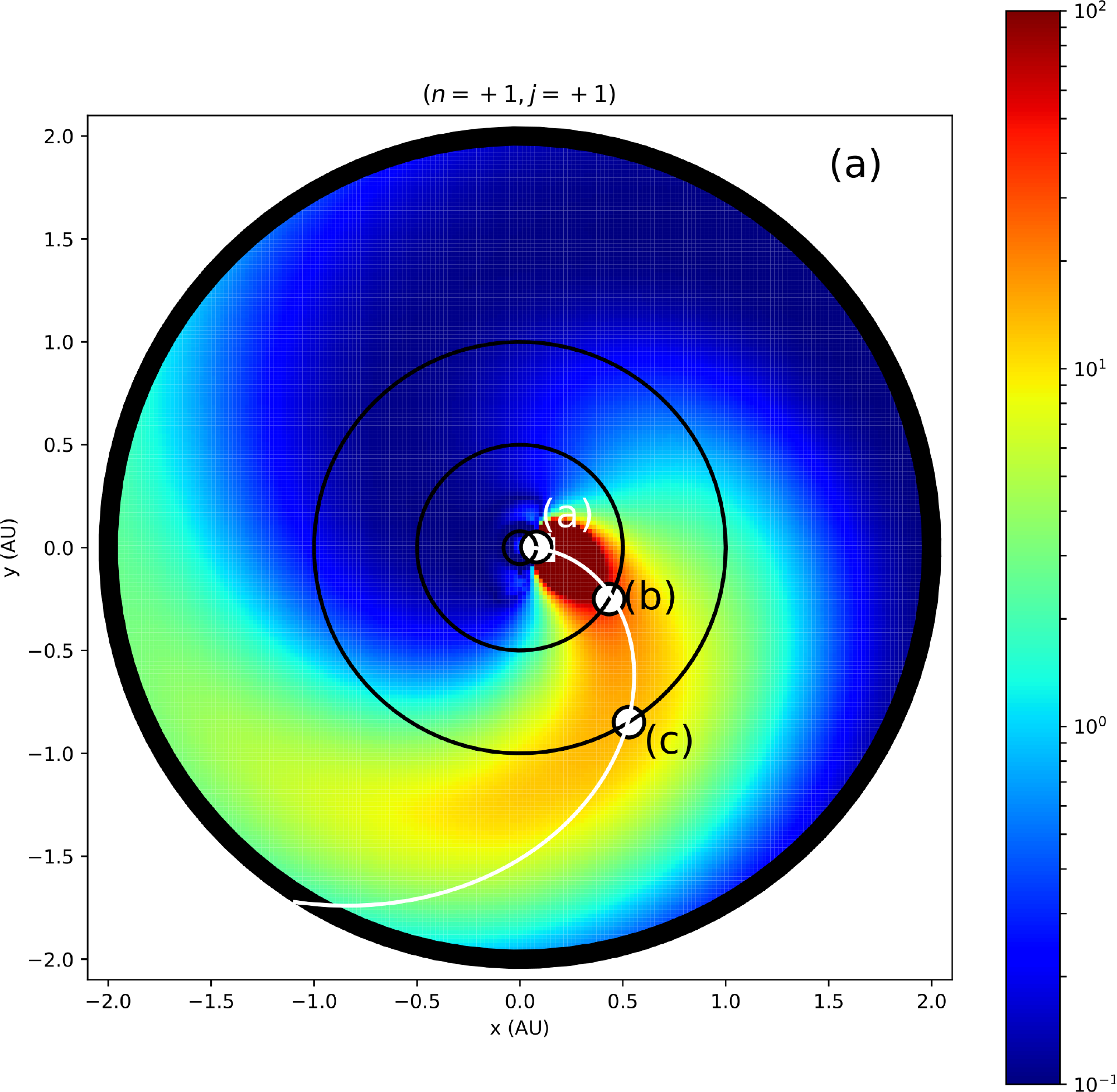}
\includegraphics[width=85mm]{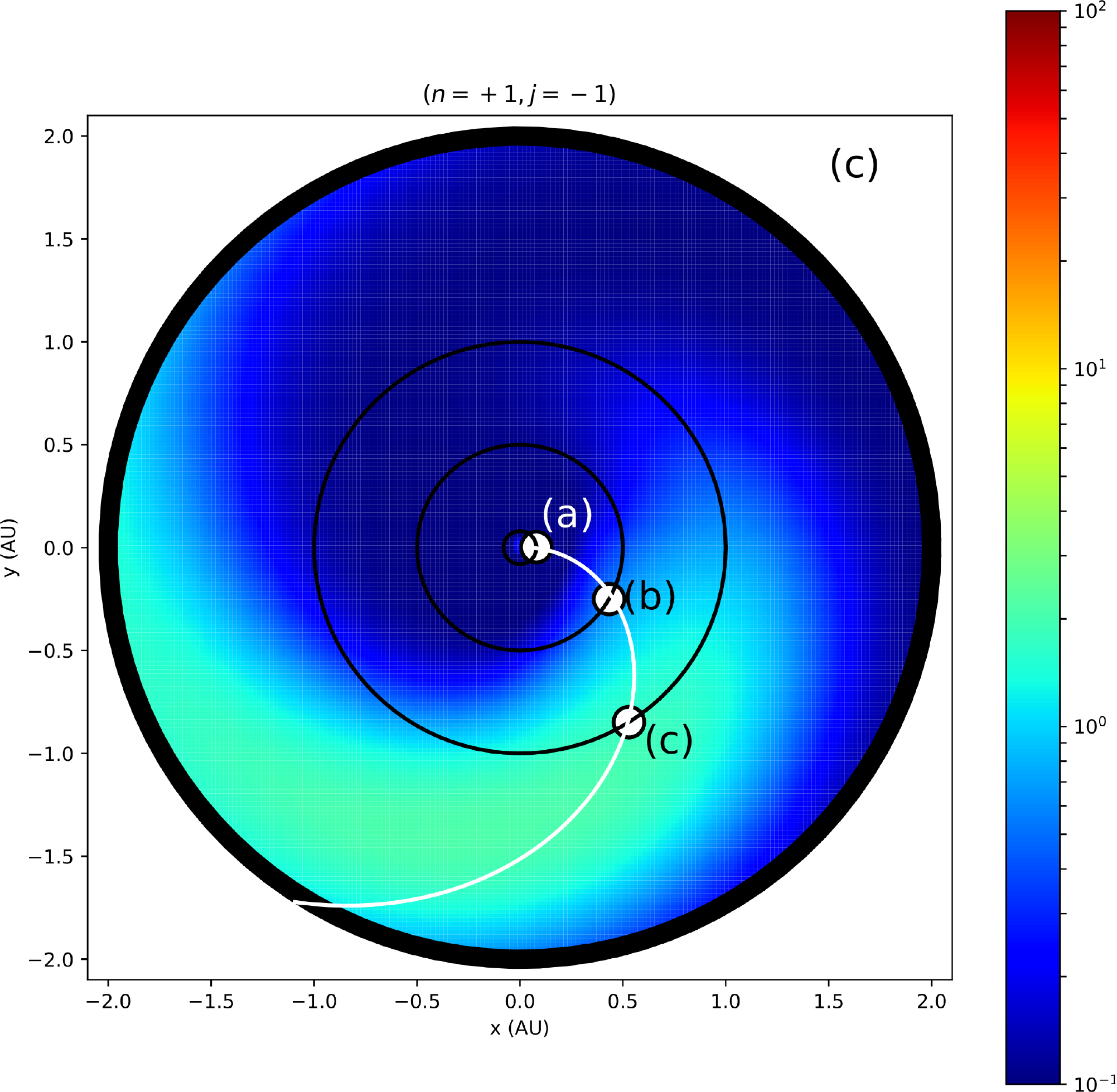}\\
\includegraphics[width=85mm]{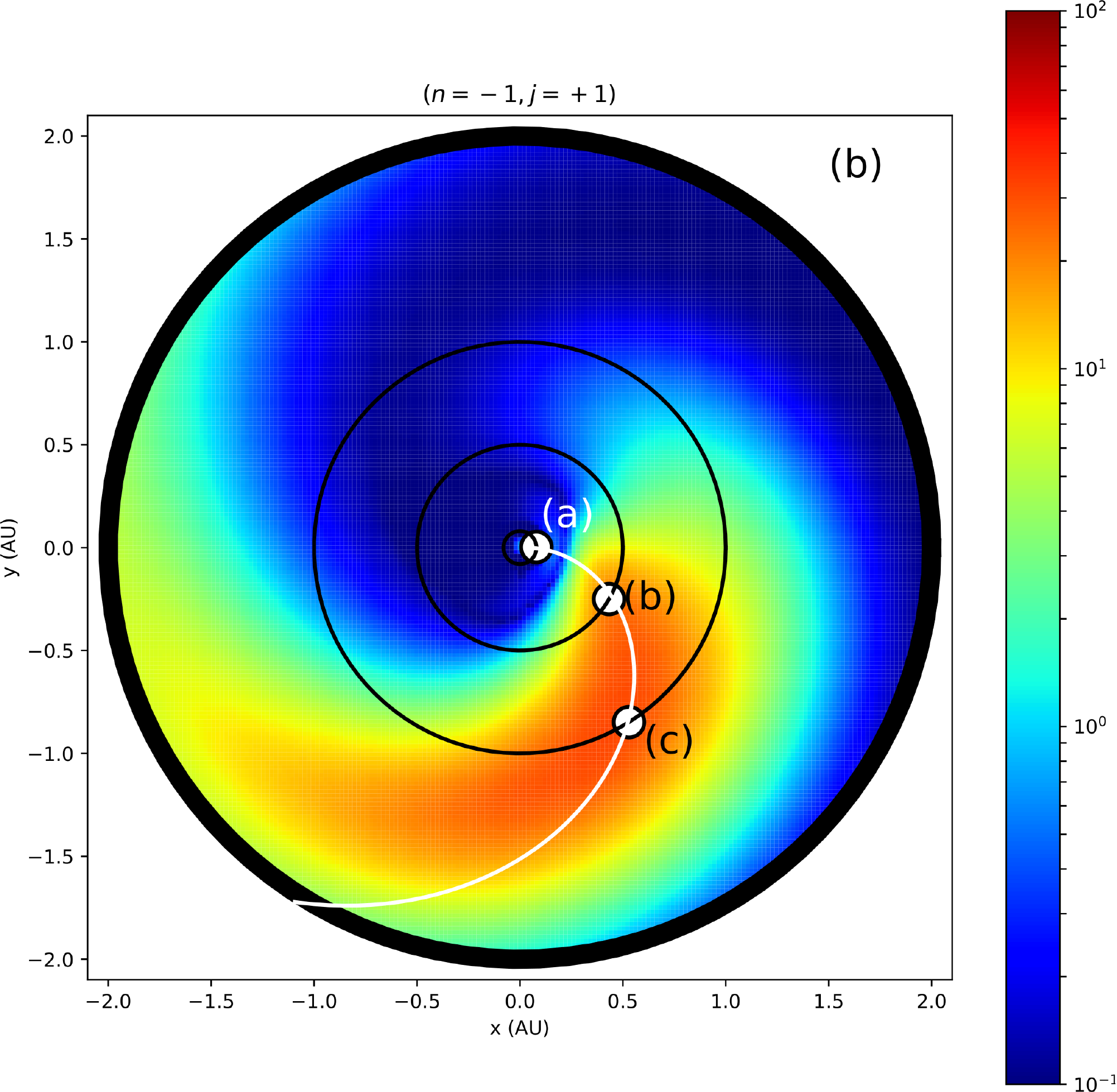}
\includegraphics[width=85mm]{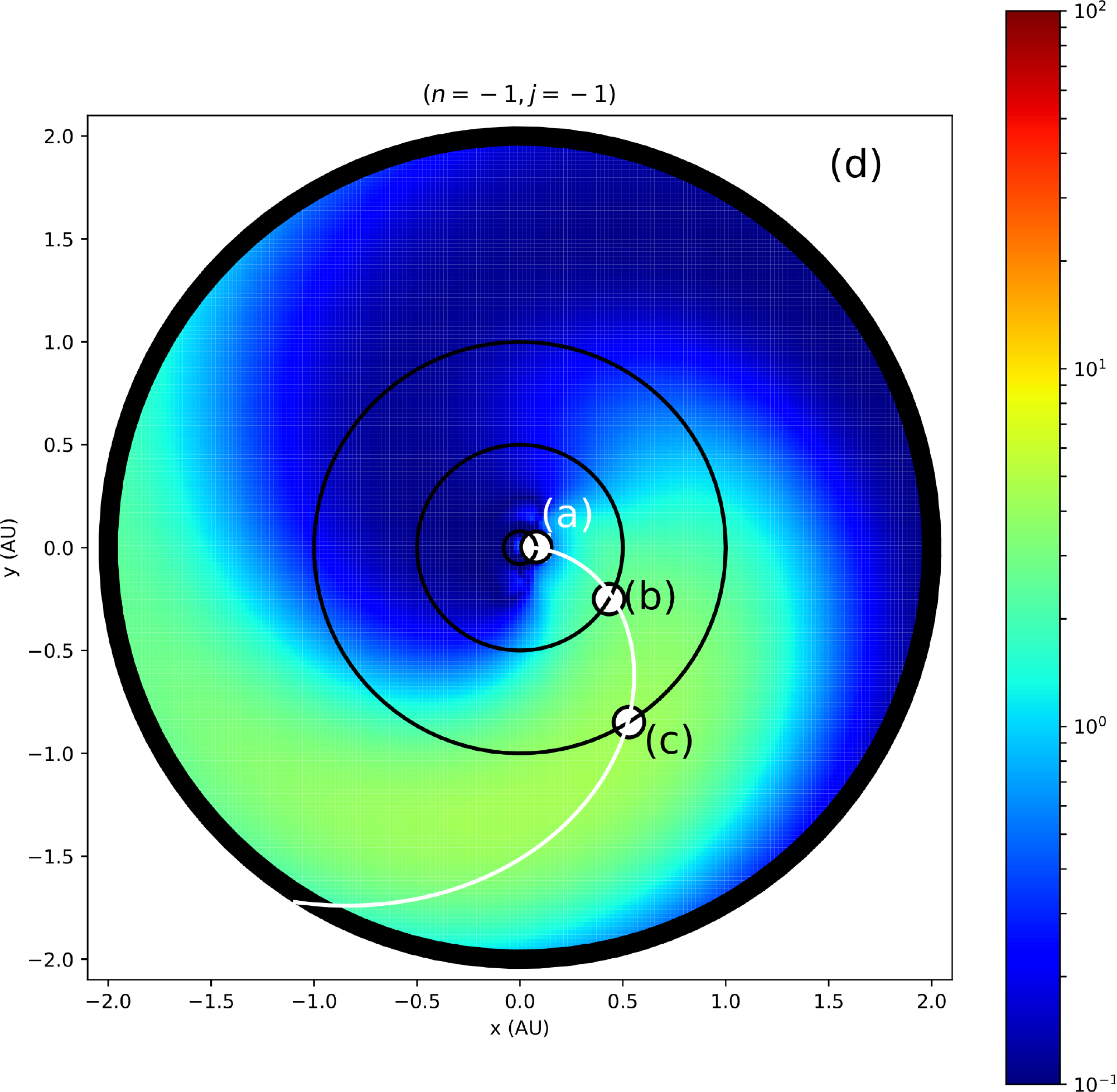}
\caption{Contour plot showing where each wavemode was most affected by the streaming electrons. See the text for more details. \label{fig:contour_plot_different_waves}}
\end{center}
\end{figure*}

\begin{figure}
\begin{center}
\includegraphics[width=80mm]{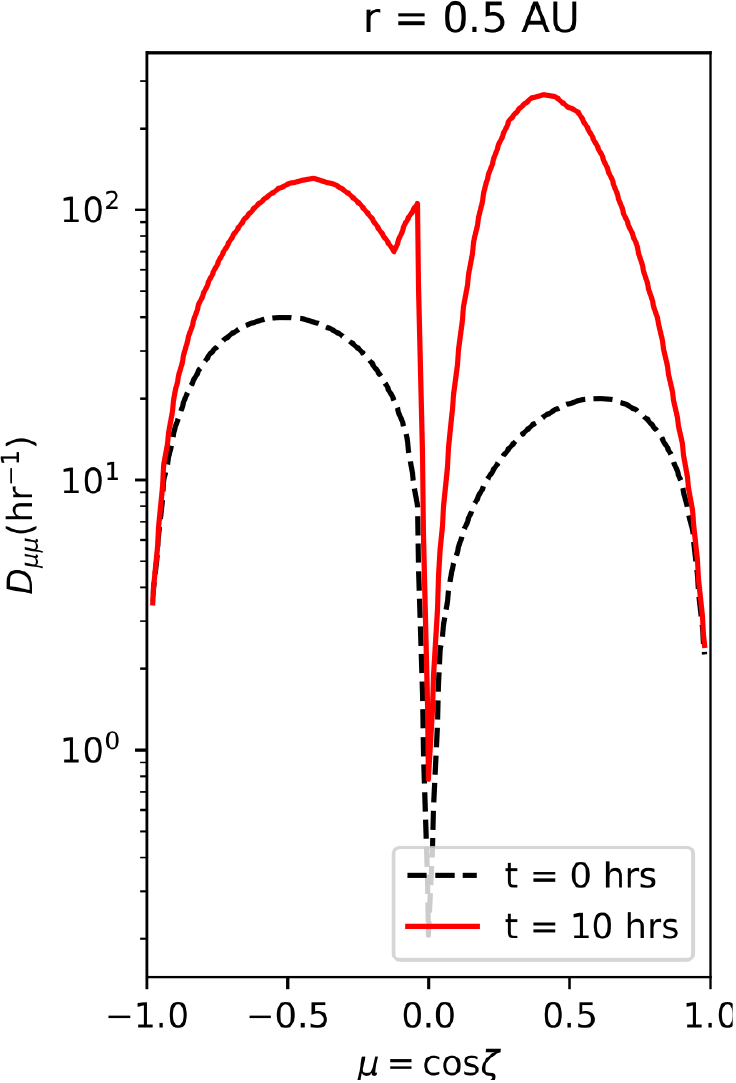}
\caption{Comparing the pitch-angle diffusion coefficient, at $r=0.5$ AU, before (dashed black line) and after (solid red line) the SEP event was simulated. \label{fig:duu_after}}
\end{center}
\end{figure}

The left panel of Fig. \ref{fig:turb_spectra_after_event} shows the pre-event energy spectra of the four waves components of the slab turbulence spectrum at a radial position of $r=0.5$ AU. The right panel shows the same spectra after the SEP events passed over the virtual spacecraft. As expected, forward propagating wavemodes were amplified, while backward propagating waves were damped. The forward propagating LH polarized modes is most heavily affected as a beamed electron population will resonate most effectively with these wavemodes. Interestingly, for 100 keV electrons under consideration here, the wave growth and/or damping occurs at the higher end of the inertial range, or possibly near the onset of the dissipation range, of the turbulence spectrum. Newer spacecraft definitely have the possibility to resolve this part of the turbulence spectrum, and results such as shown on the right panel of Fig. \ref{fig:turb_spectra_after_event} should be observable.

To quantity and illustrate where in space most of the wave modification takes place, we calculate the so-called ``residual turbulence energy" for each wave as

\begin{equation}
\Delta B_{n,j} = \left| \int_{k^{(1)}_{||}}^{k^{(2)}_{||}} \left( g^{\mathrm{slab}}(k_{||},t_1,n,j) - g^{\mathrm{slab}}(k_{||},t_2,n,j)  \right) \right|, 
\end{equation}

where $k^{(1)}_{||}$ and $k^{(2)}_{||}$ is the range of the wavenumber space under consideration (see Fig. \ref{fig:turb_spectra_after_event}), $g^{\mathrm{slab}}(k_{||},t,n,j)$ the energy spectra of the different wavemodes and $t_1 = 0$ hrs and $t_2 = 10$ hrs the start and end of the simulation interval. This quantity is shown, for each wavemode separately, as a contour plot in the ecliptic plane of the heliosphere in Fig. \ref{fig:contour_plot_different_waves}. To some extent these maps mimic the results of the differential intensity contour (see Fig. \ref{fig:intensity_contour}). This is expected as higher fluxes generally lead to more wave modification. The details are, however, much more complex, with the wave growth dependent on e.g. $D_{\mu \mu}$, the anisotropic part of the SEP distribution, and the resonance functions. As an illustration, note that the most efficient wave growth for LH polarized, forward moving waves is close to the Sun (see panel a), while most effective wave growth for RH-polarized, forward moving waves are closer to Earth orbit.

The pitch-angle diffusion coefficient, as calculated before and after the SEP event, is shown in Fig. \ref{fig:duu_after}. The change in $D_{\mu \mu}$ is pitch-angle dependent with the biggest changes occurring at intermediate pitch-angle values, where the level of scattering increased by more than an order of magnitude. This is very large, considering that only the forward propagating waves are amplified. The kink in the calculated $D_{\mu \mu}$ after the event is due to our rather coarse pitch-angle grid used in the model. See also the zig-zags present in the solutions presented in Fig. \ref{fig:turb_spectra_after_event}.

\begin{figure}
\begin{center}
\includegraphics[width=80mm]{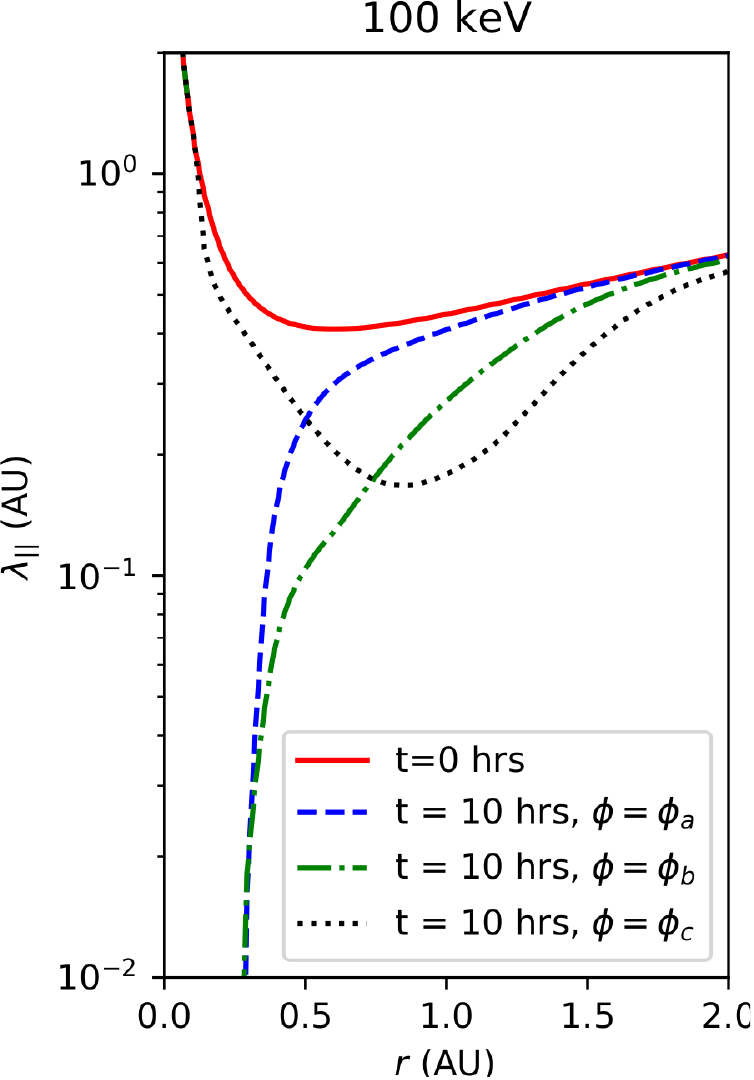}
\caption{The calculated $\lambda_{||}$, shown as a function of radial distance, at different longitudes before and after the simulated SEP event. \label{fig:lambdas_after}}
\end{center}
\end{figure}

To quantify the enhance level of scattering, we calculate $\lambda_{||}$, at different longitudes, after the simulated SEP events. The results are shown in Fig. \ref{fig:lambdas_after}. Note that the different azimuthal positions co-inside with the different virtual spacecraft placed in the computational domain (see again Fig. \ref{fig:intensity_contour}). The effect of wave growth on $\lambda_{||}$ is dramatic; near the Sun $\lambda_{||}$ decreases by at least two orders of magnitude. This decrease, however, depends very strongly on magnetic connectivity to the SEP source with an observer $\sim 45^{\circ}$ away from the source unlikely to see any larger effects. The effects of wave growth also diminishes away from the Sun, with only moderate effects apparent at Earth. 

\begin{figure}
\begin{center}
\includegraphics[width=80mm]{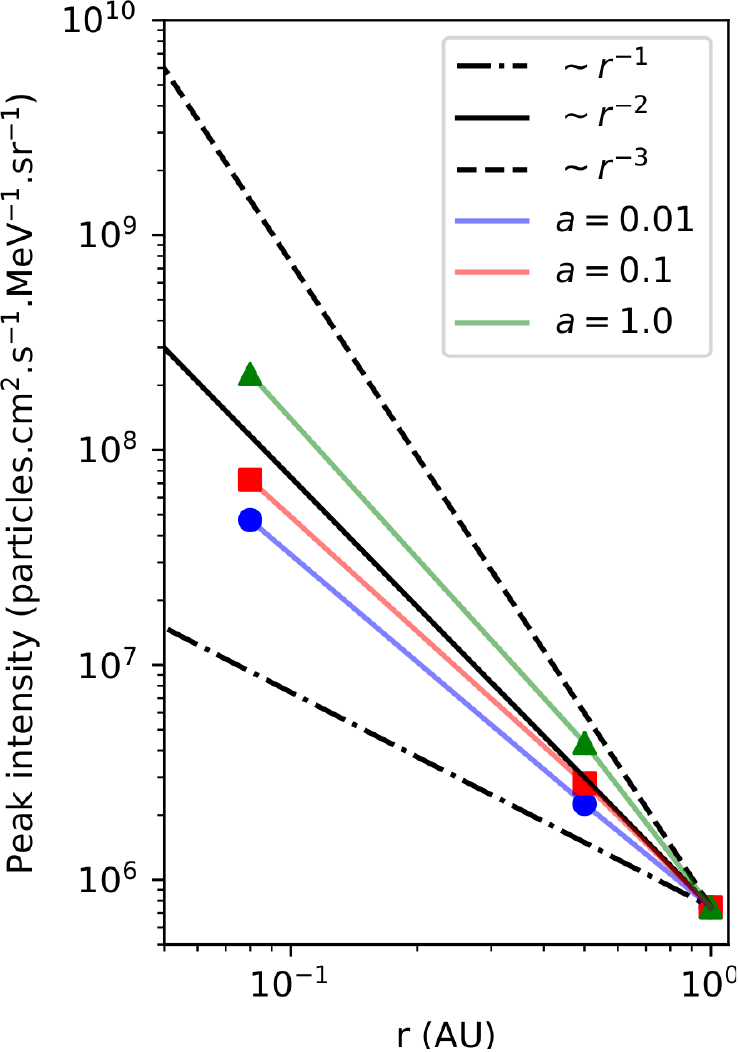}
\caption{The calculated peak-intensity, along an optimally connected magnetic field line, as a function of radius, for various assumptions of the parameter $a \in [0,1]$ that determines the strength of perpendicular diffusion. Note that all simulations are normalized to the same value at Earth. \label{fig:perp_diff}}
\end{center}
\end{figure}

\subsection{The effects of perpendicular diffusion}

Perpendicular diffusion may also influence the calculated wave growth-rates indirectly by controlling the intensity of SEPs near the Sun: with more effective perpendicular diffusion in the model, it is expected that the SEP particles will spread out more effectively, in terms of longitude, with less particles reaching Earth. However, as we normalize the intensities to Earth values, more particle spreading means that the initial SEP distribution specified near the Sun must be higher. Naively, this argument suggests that more efficient perpendicular diffusion will lead to more efficient wave-growth.

To test this argument, we have performed simulations with various choices of $a \in [0,1]$ leading to different levels of perpendicular diffusion. The resulting peak intensities are normalized to the same value at Earth, and Fig. \ref{fig:perp_diff} shows the radial dependence thereof. Note that we calculate the peak intensity along the magnetic field line that connect each point to the SEP source. Referring to Fig. \ref{fig:perp_diff}, we see that the behaviour is well described by a $\sim 1/r^2$ dependence, and that the peak intensity near the source only changes by a factor of $\sim 2$ between the extreme values of $a = 0.01$ and $a=1$. 

Interestingly, our model results compare well with these of \citet{heetal2017} who found a $r^{-1.7}$ dependence using a much more involved 3D modelling approach.

\section{Discussion}

In this paper we have calculated diffusion coefficients for 100 keV electrons in the inner heliosphere based on observed turbulence spectra. Effects such as non-zero magnetic and cross helicities, which are usually neglected, have been included and it was shown shown to have large effects near the Sun. We have also included dynamical effects in the standard FLRW perpendicular diffusion coefficient which may have large effects for low energy particles. 

Using these theoretically motivated transport coefficients, we have included, for the first time, wave generation by streaming SEP electrons. Our results show that, for the biggest events, SEP electrons can significantly grow forward propagating waves (especially left-handed polarized waves near the Sun) {Our results therefore confirm, to some extent, the observations of \citet{aguedalario}}. These effect diminish towards Earth and are not expected to observable outside of 0.5 AU and when the observer is not magnetically well connected to the flaring region. However, the upcoming {\it Parker Solar Probe} and {\it Solar Orbiter} spacecraft should be able to observe these amplified levels of turbulence in the coming years.

{Our model of wave growth is by no means complete; a full spectrum of particles needs to be injected into interplanetary space, with wave cascading and dissipation processes treated more self consistently. Wave attenuation effects will also have to be included in future. Such a model is however not computationally viable at the moment.} We are also not able, at present, to confirm whether a so-called streaming limit \citep[e.g.][]{reamesng2010} is present in the electron simulations. {However, we expect that the simulated temporal evolution of the SEP electron intensity at e.g. Earth will be greatly affected by wave growth for the largest SEP events given the large changes in $D_{\mu \mu}$. Interestingly, we note that the enhanced turbulence levels (and correspondingly small $\lambda_{||}$) near the Sun can strongly enhance the diffusive shock acceleration rate in this region, possibly boosting the level of energetic electrons reaching Earth when a shock from a coronal mass ejection is present.}

We have modelled the peak intensity along the MHF line of optimal magnetic connectivity. Our calculations compare well with previous 3D modelling results and the observations of \citet{larioetal2013}. Moreover, it was shown that the level of perpendicular diffusion can influence the wave growth rate by changing the SEP intensity near the Sun (as normalized to constant levels at Earth). We again emphasize that perpendicular diffusion is an important transport process, as also illustrated by \citet{wolfgang}, \citet{he2015}, \citet{Laitinenetal2016} and \citet{zhaozhang2018}, amongst others, and should be included in SEP transport models.

We are confident that our approach of using theoretically derived coefficients, combined with turbulence quantities, consistent with the limited available observations, can produce realistic SEP intensities in the inner heliosphere. The approach can be refined with more measurement, but is the most effective way to progress towards a predictive model and understanding of the fundamental processes that couple particle scattering to solar wind fluctuations. Moreover, most of these turbulence quantities can be more constrained by the upcoming {\it Parker Solar Probe} and {\it Solar Orbiter} spacecraft.

\acknowledgments

{This work is based on research supported in part by the National Research Foundation (NRF) of South Africa (grant no. 106049). Opinions expressed and conclusions arrived at are those of the authors and are not necessarily to be attributed to the NRF. RDS acknowledges partial financial support from the Fulbright Visiting Scholar Program.}

\appendix

\section{Calculation of the growth/damping rate}
\label{Sec:appendix_1}

Below we show the detailed calculation of the wave growth/damping rate. The derivation is loosely based on work presented by e.g. \citet{ngreames1994} and \citet{vainio2003}, amongst others.

The change in wave energy during a scattering event can be approximated as $\Delta E_w = - j V_A p \Delta \mu$, with the total contribution from all particles in the distribution giving the total change in the energy density of the waves

\begin{equation}
\frac{d U_w}{dt}= -j \int d^3 p  V_A p \frac{\langle \Delta \mu \rangle}{\Delta t} f (\vec{r}, \vec{p}, t).
\end{equation}

Assuming pitch-angle scattering to be the only process changing the distribution, we may estimate this change as

\begin{equation}
\frac{\partial f}{\partial t} = \frac{\partial}{\partial \mu} \left(D_{\mu\mu}  \frac{\partial f}{\partial \mu} \right) \approx - \frac{\partial}{\partial \mu} \left( \frac{\langle \Delta \mu \rangle}{\Delta t} f \right),
\end{equation}

where the second part of the expression follows from the standard definition of the Fokker-Planck equations, resulting in

\begin{equation}
\frac{\langle \Delta \mu \rangle}{\Delta t} f \approx - D_{\mu \mu} \frac{\partial f}{\partial \mu},
\end{equation}

and hence that the average change in wave energy density is given by

\begin{equation}
\frac{d U_w}{dt}= j V_A \int d^3 p  p D_{\mu\mu}  \frac{\partial f (\vec{r}, \vec{p}, t)}{\partial \mu}.
\end{equation}

Assuming $f$ to be gyro-tropic, we find

\begin{equation}
\frac{d U_w}{dt}=  2 \pi j  V_A  \int d\mu \int dp p^3 D_{\mu\mu}  \frac{\partial f (\vec{r}, p, \mu, t)}{\partial \mu}.
\end{equation}

In order to make wave-generation implementable in our numerical model, we have to assume a mono-energetic SEP distribution, i.e. $f (\vec{r}, p, \mu, t) = p_0 f(\vec{r}, \mu, t) \delta (p - p_0)$, so that

\begin{equation}
\label{Eq:wave_growth_1}
\frac{d U_w}{dt}=  2 \pi j  V_A  p_0^4  \int d\mu  D_{\mu\mu}   \frac{\partial f (\vec{r}, p, \mu, t)}{\partial \mu}.
\end{equation}

If the amplified waves are Alfv{\'e}nic, both the energy density of the kinetic and magnetic contributions change (with an equipartition for non-dispersive Alfv{\'e}n waves), resulting in

\begin{equation}
U_w = \delta U_{\mathrm{magnetic}} + \delta U_{\mathrm{kinetic}} = 2 \delta U_{\mathrm{magnetic}} = \frac{\delta B_{\mathrm{slab}}^2}{4 \pi}.
\end{equation}

By using Eq. \ref{Eq:def_db_slab}, the change of energy density can be evaluated as a change in wave power,

\begin{equation}
\label{Eq:wave_energy_derivative}
\frac{dU_w}{dt} = \frac{1}{2} \int_0^{\infty} dk_{||} \sum_{n = \pm 1} \sum_{j=\pm1} \frac{d g^{\mathrm{slab}}_{n,j}}{d t}.
\end{equation}

Combining Eqs. \ref{Eq:wave_growth_1} and \ref{Eq:wave_energy_derivative}, we get

\begin{equation}
\label{Eq:wave+growth_2}
\int_0^{\infty} dk_{||} \sum_{n = \pm 1} \sum_{j=\pm1} \frac{d g_{slab}^{n,j}}{d t} -  4 \pi j  V_A  p_0^4  \int d\mu  D_{\mu\mu}   \frac{\partial f (\vec{r}, p, \mu, t)}{\partial \mu} = 0.
\end{equation}

and we now need to estimate the pitch-angle diffusion coefficient, $D_{\mu \mu}$. In order to get a tractable expression, we use the {\it negligible damping} version, given by Eqs. \ref{Eq:non_damped_1} -- \ref{Eq:non_damped_3}, to obtain

\begin{equation}
\int_0^{\infty} dk_{||} \sum_{n = \pm 1} \sum_{j=\pm1} g^{\mathrm{slab}}_{n,j} \left[ \frac{1}{g^{\mathrm{slab}}_{n,j}} \frac{d g^{\mathrm{slab}}_{n,j}}{d t} -  \frac{4 \pi^3 j p_0^4 V_A \Omega^2}{B_0^2}  \int d\mu  (1 - \mu^2)  \delta \left( v \mu j k_{||} - \omega + n\Omega \right)  \frac{\partial f }{\partial \mu} \right]= 0.
\end{equation}

As this expression should be true for each independent wavemode, we can use Eq. \ref{Eq:define_growth_rate} to find the growth/damping rate for the different wavemodes as

\begin{equation}
2 \gamma_{n,j} =  \frac{4 \pi^3 j p_0^4 V_A \Omega^2}{B_0^2}  \int d\mu  (1 - \mu^2)  \delta \left( v \mu j k_{||} - \omega + n\Omega \right)  \frac{\partial f }{\partial \mu} .
\end{equation}

Manipulating the $\delta$-function gives

\begin{equation}
\label{Eq:final_wave_growth}
\gamma_{n,j} =  j \frac{\pi^2}{4} \frac{p_0^4}{k^{\mathrm{res}}_{||}} \frac{\Omega^2}{  \mathcal{B}} \frac{V_A}{v}   (1 - \mu_{\mathrm{res}}^2)  \left. \frac{\partial f }{\partial \mu} \right|_{\mu_{\mathrm{res}}} ,
\end{equation}

where the resonant pitch-angle is

\begin{equation}
\mu_{\mathrm{res}} = \frac{\omega - n\Omega}{jvk^{\mathrm{res}}_{||}},
\end{equation}

and the energy is the background magnetic field is

\begin{equation}
\mathcal{B} = \frac{B_0^2}{8 \pi}.
\end{equation}

\section{Plasma dispersion relations}
\label{Sec:appendix_2}

For the sake of completeness, we list the dispersion relations for circularly polarized waves in the cold plasma limit, that is, assuming $T_\mathrm{e} = T_\mathrm{p} = 0$. These relations are given by \citep[see, amongst other,][]{stix}

\begin{equation}
\label{Eq:LH_dispersion}
k_{||,\mathrm{LH}}^2 \approx \frac{\omega^2}{V_A^2} \left\{  \frac{\Omega_p}{\Omega_p - \omega}  \right\}
\end{equation}

and

\begin{equation}
\label{Eq:RH_dispersion}
k_{||,\mathrm{RH}}^2 \approx \frac{\omega^2}{c^2} \left\{  \frac{|\Omega_e|}{|\Omega_e| - \omega} + \frac{\omega_e^2}{(|\Omega_e| - \omega)(\Omega_p + \omega)}  \right\},
\end{equation}

where $k_{||}$ is the parallel wavenumber, $\omega$ the frequency in the {\it bulk (flow) plasma frame}, $c$ the speed of light and $V_A$ the Alfv\'en speed. $\Omega_p$ and $|\Omega_e|$ are the proton and electron cyclotron frequencies, given as

\begin{equation}
|\Omega_e| = \frac{|e|B}{m_e} = \Lambda \Omega_p
\end{equation}

where $\Lambda = m_p/m_e$ is the ratio between proton and electron masses, $e$ the elementary charge and $B$ the mean background magnetic field, while the different plasma frequencies are given as

\begin{equation}
\omega_e = \sqrt{\frac{n_e e^2}{m_e \epsilon_0}} = \omega_p \sqrt{\Lambda},
\end{equation}

with $n_e$ the electron number density and $\epsilon_0$ the permittivity of free space. Note that we assume quasi-neutrality with $n_e \approx n_p$. In terms of these frequencies, $V_A$ is

\begin{equation}
V_A = c \frac{\Omega_p}{\omega_p}.
\end{equation}

In deriving Eqs. \ref{Eq:LH_dispersion} and \ref{Eq:RH_dispersion} it was assumed that $\omega \ll \omega_p \ll \omega_e$. The dispersion relations are shown in Fig. \ref{fig:dispersion_diagram} as the solid blue lines.

In the cold plasma limit, the waves are strongly damped near either the proton or electron cyclotron frequency. However, when a {\it warm} plasma is considered (such as the solar wind), the thermal motion of the plasma particles, with a speed of

\begin{equation}
v_{\mathrm{th}}^{p,e} = \sqrt{\frac{2kT_{p,e}}{m_{p,e}}},
\end{equation}

for either protons or electrons, need to be considered. In Eq. \ref{Eq:k_d_onset} this was done in a rather {\it ad-hoc} way by relying on the results of \citet{cedricetal2017}. Indeed, this approach was also shown by \cite{ES2018} to compare well with observed quantities. The difficulty lies is calculating, in a tractable analytical fashion, the damping rate of MHD waves by a warm plasma. Unfortunately, this is only possible under the assumption of a small growth rate, $\gamma_{\omega} \ll \omega$ and leads to

\begin{equation}
\gamma_{\omega} = - D(\omega, k_{||}) / \left( \frac{\partial D(\omega, k_{||})}{\partial \omega} \right)
\end{equation}

and needs to be evaluated for each polarization mode separately \citep[for details, see e.g.][]{chenetal2013}. The function $D$ is related to the distribution of the particles under consideration, and for our purposes, we additionally assume no temperature anisotropy, so that

\begin{equation}
D(\omega, k_{||}) = \sqrt{\pi} \frac{\omega}{k_{||}} \sum_{p,e}  \frac{ \omega_{p,e}^2}{v^{p,e}_{\mathrm{th}}}  \exp \left[ - \left( \frac{\omega + n\Omega_{p,e}}{k_{||} v^{p,e}_{\mathrm{th}}} \right)^2 \right],
\end{equation}

and

\begin{equation}
\frac{\partial D(\omega, k_{||})}{\partial \omega} = 2 \omega - n \sum_{p,e}  \omega_{p,e}^2 \frac{\Omega_{p,e}}{\left( \omega + n \Omega_{p,e} \right)^2},
\end{equation}

where, for the latter equation, we have furthermore assumed that $v^{p,e}_{\mathrm{th}} \rightarrow 0$. The calculated damping rates, using 1 AU plasma values, are shown in Fig. \ref{fig:dispersion_diagram} as the solid red curves.




\end{document}